\documentclass[aps,twocolumn,showpacs,preprintnumbers,nofootinbib,prd,superscriptaddress,groupedaddress,10pt]{revtex4-1}

\makeatletter
\def\l@subsubsection#1#2{}
\def\l@subsubsubsection#1#2{}
\makeatother

\usepackage{graphicx,amssymb,amsmath,amsthm,amsfonts,epsfig,epsf}
\usepackage[usenames]{color}
\usepackage{epstopdf}

\usepackage{aas_macros}
\usepackage{bm}
\usepackage{dcolumn}
\usepackage[latin1]{inputenc}
\usepackage{latexsym}
\usepackage{rotating}
\usepackage{longtable}

\usepackage{enumerate}
\usepackage{tensor,multirow}
\usepackage{url}
\usepackage[linktocpage]{hyperref}

\def\nn{\nonumber}
\def\polylog{\operatorname{polylog}}

\def\be{\begin{equation}}
\def\ee{\end{equation}}
\newcommand{\beq}{\begin{eqnarray}}
\newcommand{\eeq}{\end{eqnarray}}
\newcommand{\C}{\mathcal{C}}

\def\ba{\begin{align}}
\def\ea{\end{align}}
\newcommand{\tn}{\textnormal}

\begin{document}
\title{Testing strong-field gravity with tidal Love numbers}

\author{
Vitor Cardoso,$^{1,2}$ 
Edgardo Franzin,$^{3}$
Andrea Maselli,$^4$
Paolo Pani,$^{5,1}$
Guilherme Raposo$^{1}$
}
\affiliation{${^1}$ CENTRA, Departamento de F\'{\i}sica, Instituto Superior T\'ecnico -- IST, Universidade de Lisboa -- UL,
Avenida Rovisco Pais 1, 1049 Lisboa, Portugal}
\affiliation{${^2}$ Perimeter Institute for Theoretical Physics, 31 Caroline Street North
Waterloo, Ontario N2L 2Y5, Canada}
\affiliation{${^3}$ Dipartimento di Fisica, Universit\`a di Cagliari \& Sezione INFN Cagliari, Cittadella Universitaria, 09042 Monserrato, Italy}
\affiliation{$^{4}$ Theoretical Astrophysics, Eberhard Karls University of Tuebingen, Tuebingen 72076, Germany}
\affiliation{$^{5}$ Dipartimento di Fisica, ``Sapienza'' Universit\`a di Roma \& Sezione INFN Roma1, Piazzale Aldo Moro 5, 00185, Roma, Italy}
\begin{abstract}
The tidal Love numbers (TLNs) encode the deformability of a self-gravitating object immersed in a tidal environment and depend significantly both on the object's internal structure and on the dynamics of the gravitational field. 
An intriguing result in classical general relativity is the vanishing of the TLNs of black holes. We extend this result in three ways, aiming at testing the nature of compact objects: (i) we compute the TLNs of exotic compact objects, including different families of boson stars, gravastars, wormholes, and other toy models for quantum corrections at the horizon scale. In the black-hole limit, we find a universal logarithmic dependence of the TLNs on the location of the surface; (ii) we compute the TLNs of black holes beyond vacuum general relativity, including Einstein-Maxwell, Brans-Dicke and Chern-Simons gravity;
(iii) We assess the ability of present and future gravitational-wave detectors to measure the TLNs of these objects,
including the first analysis of TLNs with LISA. 
Both LIGO, ET and LISA can impose interesting constraints on boson stars, while LISA is able to probe even extremely compact objects. 
We argue that the TLNs provide a smoking gun of new physics at the horizon scale, and that future gravitational-wave measurements of the TLNs in a binary inspiral provide a novel way to test black holes and general relativity in the strong-field regime.
\end{abstract}

\maketitle

\tableofcontents

\section{Introduction}

Tidal interactions play a fundamental role in astrophysics across a broad range of scales, from stellar objects like ordinary stars and neutron stars (NSs) to large celestial systems such as galaxies. Several astrophysical structures (e.g., binaries and tidal tails~\cite{Toomre:1972vt,1977ApJ...213..183P}) are consequences of tidal interactions. Tidal effects can be particularly strong and important in the regime that characterizes compact objects, giving rise to extreme phenomena such as tidal disruptions. 

The deformability of a self-gravitating object immersed in an external tidal field is measured in terms of its tidal Love numbers~(TLNs)~\cite{Murraybook,PoissonWill}. These leave a detectable imprint in the gravitational-wave (GW) signal emitted by a neutron-star binary in the late stages of its orbital evolution~\cite{Flanagan:2007ix,Hinderer:2007mb,Hinderer:2016eia}.
So far, a relativistic extension~\cite{Hinderer:2007mb,Binnington:2009bb,Damour:2009vw} of the Newtonian theory of tidal deformability has been mostly motivated by the prospect of measuring the TLNs of NSs through GW detections and, in turn, understanding the behavior of matter at supranuclear densities~\cite{Lattimer:2004pg,Hinderer:2009ca,Postnikov:2010yn,Vines:2011ud,Damour:2012yf,DelPozzo:2013ala,Maselli:2013rza}.
The scope of this paper is to show that tidal effects can also be used to explore more fundamental questions related to the nature of compact objects and the behavior of gravity in the strong-field regime (for a previous related study in the context of aLIGO binaries, cf.\ Ref.~\cite{Wade:2013hoa}).\footnote{A related, independent work dealing with tidal effects for boson stars, conducted simultaneously to ours, is due to appear soon~\cite{Buonanno:2017}.}

An intriguing result in classical general relativity~(GR) is the fact that the TLNs of a black hole (BH) are precisely zero. This property has been originally demonstrated for small tidal deformations of a Schwarzschild BH~\cite{Binnington:2009bb,Damour:2009vw,Fang:2005qq} and has been recently extended to arbitrarily strong tidal fields~\cite{Gurlebeck:2015xpa} and to the spinning case~\cite{Poisson:2014gka,Pani:2015hfa,Landry:2015zfa}, at least in the axisymmetric case to quadratic order in the spin~\cite{Pani:2015hfa} and generically to linear order in the spin~\cite{Landry:2015zfa}.

\subsection{The naturalness problem}
The precise cancellation of the TLNs of BHs within Einstein's theory poses a problem of ``naturalness'' in classical GR~\cite{Porto:2016pyg,Porto:2016zng,Rothstein:simons}, one that can be argued to be as puzzling as the strong CP and the hierarchy problem in particle physics, or as the cosmological constant problem. The resolution of this issue in BH physics could lead to --~testable, since they would be encoded in GW data~--smoking-gun effects of new physics. 

This question can be solved in at least two (related) ways, which we explore here. If new physics sets in, for example through unexpectedly large quantum back-reaction or changes in the equation of state, BHs might simply not be formed, 
avoiding this and other problems (such as the information loss puzzle~\cite{Mathur:2009hf}) altogether.
Instead, other objects might be the end product of gravitational collapse.
These ``exotic compact objects'' (ECOs) include boson stars (BSs)~\cite{Schunck:2003kk,Liebling:2012fv,Brito:2015pxa,Brito:2015yfh,Grandclement:2016eng}, gravastars~\cite{Mazur:2001fv,Visser:2003ge}, wormholes~\cite{visser1995lorentzian}, 
and various toy models describing quantum corrections at the horizon scale, like superspinars~\cite{Gimon:2007ur}, fuzzballs~\cite{Skenderis:2008qn}, ``2-2 holes''~\cite{Holdom:2016nek} and others~\cite{Saravani:2012is,Giddings:2014ova,Abedi:2016hgu}.
ECOs might be formed from the collapse of exotic fields or by quantum effects at the horizon scale, and represent the prototypical example of exotic GW sources~\cite{Pani:2009ss,Macedo:2013jja,Giudice:2016zpa} which might be searched for with ground- or space-based detectors.

Alternatively, GR might not be a good description of the geometry close to horizons. BHs other than Kerr arise in theories beyond GR which are motivated by both theoretical arguments and by alternative solutions to the dark matter and the dark energy problems (for recent reviews on strong-field tests of gravity in the context of GW astronomy, see Refs.~\cite{Yunes:2013dva,Berti:2015itd}).
Arguably, the simplest BHs arise in Einstein-Maxwell theory and are described by the Reissner-Nordstr\"om solution. Although astrophysical BHs are expected to be electrically neutral~\cite{Barausse:2014tra}, Reissner-Nordstr\"om BHs can be studied as a proxy of BHs beyond vacuum GR and could also emerge naturally in models of minicharged dark matter and dark photons~\cite{Cardoso:2016olt}.
In several scalar-tensor theories, BHs are uniquely described by the Kerr solution, as in GR~\cite{hawking1972,Sotiriou:2011dz}. However, these theories introduce a scalar degree of freedom (nonminimally) coupled to gravity and the response of BHs to external perturbations is generically richer~\cite{Barausse:2008xv}. In theories with several or with complex bosons, hairy BH solutions might exist that can be seen as BSs with a BH at the center~\cite{Herdeiro:2014goa,Herdeiro:2016tmi}. These solutions could be the endpoint of the superradiant instability of the Kerr geometry, and may even describe metastable states when a single real field is present~\cite{Brito:2015oca,Brito:2014wla}.
Finally, in quadratic theories of gravity the Einstein-Hilbert action is considered as the first term of a possibly infinite expansion containing all curvature invariants, as predicted by some scenarios related to string theory and to loop quantum gravity~\cite{Berti:2015itd}. To leading order in the curvature corrections, stationary BHs in these theories belong to only two families~\cite{Yunes:2011we,Pani:2011gy}, usually dubbed the Einstein-dilaton-Gauss-Bonnet solution~\cite{Mignemi:1992nt,Kanti:1995vq,Pani:2009wy} and the Chern-Simons solution~\cite{Yunes:2009hc,Alexander:2009tp}.

\subsection{Quantifying the existence of horizons}
The observational determination of the tidal properties of compact objects also has a bearing on another fundamental question:
do event horizons exist, and how can we quantify their existence in GW data?

It was recently shown that ultracompact horizonless geometries are expected to mimic very well the last stages of coalescence of two BHs, when they merge to form a single distorted BH, ringing down to its final Kerr geometry~\cite{Cardoso:2016oxy,Cardoso:2016rao}. In this scenario, horizonless geometries would show up as {\it echoes} in the gravitational waveforms at very late times. The exclusion of echoes up to some instant $t$ after the merger rules out structure in the spacetime down to a region
$r/r_+-1\sim \exp(-t/r_+)$, with $r_+$ being the Schwarzschild radius of the spacetime. Thus, more sensitive detectors will probe regions closer and closer to the horizon.

The above picture refers to the final, postmerger state. The understanding of the initial state can start by inferring from the inspiral signal the imprints of the structure of the inspiralling objects. Putative structures will show up in the way each of these objects reacts to the gravitational field created by the other, in other words, by their TLNs. As we will show, the TLNs of all ECOs vanish in the BH limit, logarithmically. Thus, observational upper bounds on the TLNs can be converted into constraints on the compactness of the inspiralling objects.

Throughout this work, we use $G=c=1$ units and denote the Planck length by $\ell_P\approx 1.6\times 10^{-33}\,{\rm cm}$.
\subsection{Executive summary}
For the busy reader, in this section we summarize our main results; possible extensions are discussed in Sec.~\ref{sec:conclusion}.

We focus on spherically symmetric, static background geometries, and compute the TLNs under the assumption that the only surviving tide at 
large distances is gravitational. 
In this setting, the TLNs can be divided into two classes according to their parity: an electric- or polar-type, and a magnetic or axial-type, and each of these sectors can in turn be expanded into a set of multipoles labeled by an integer $l$.

\begin{figure*}[th]
\centering
\includegraphics[width=0.258\textwidth]{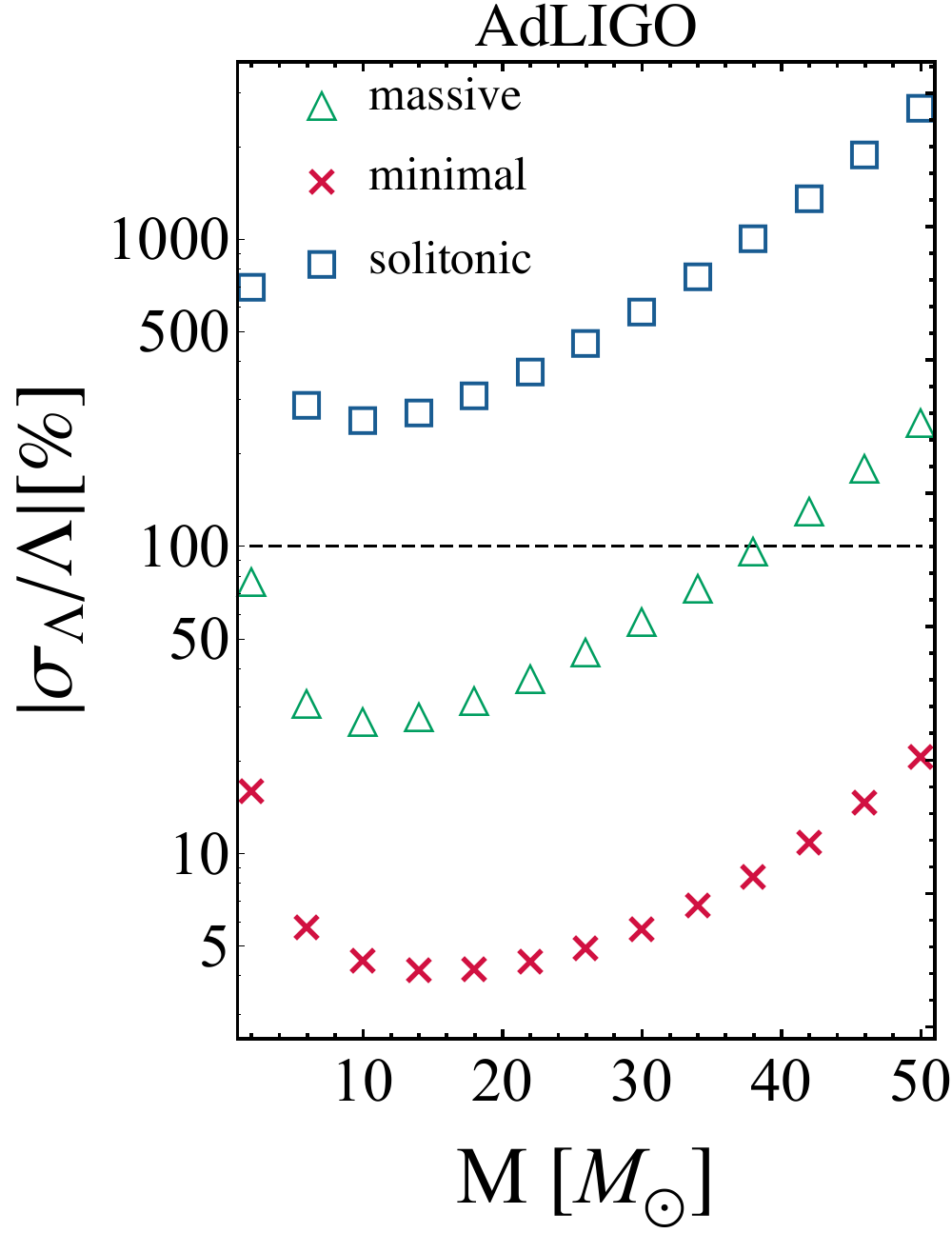}
\includegraphics[width=0.248\textwidth]{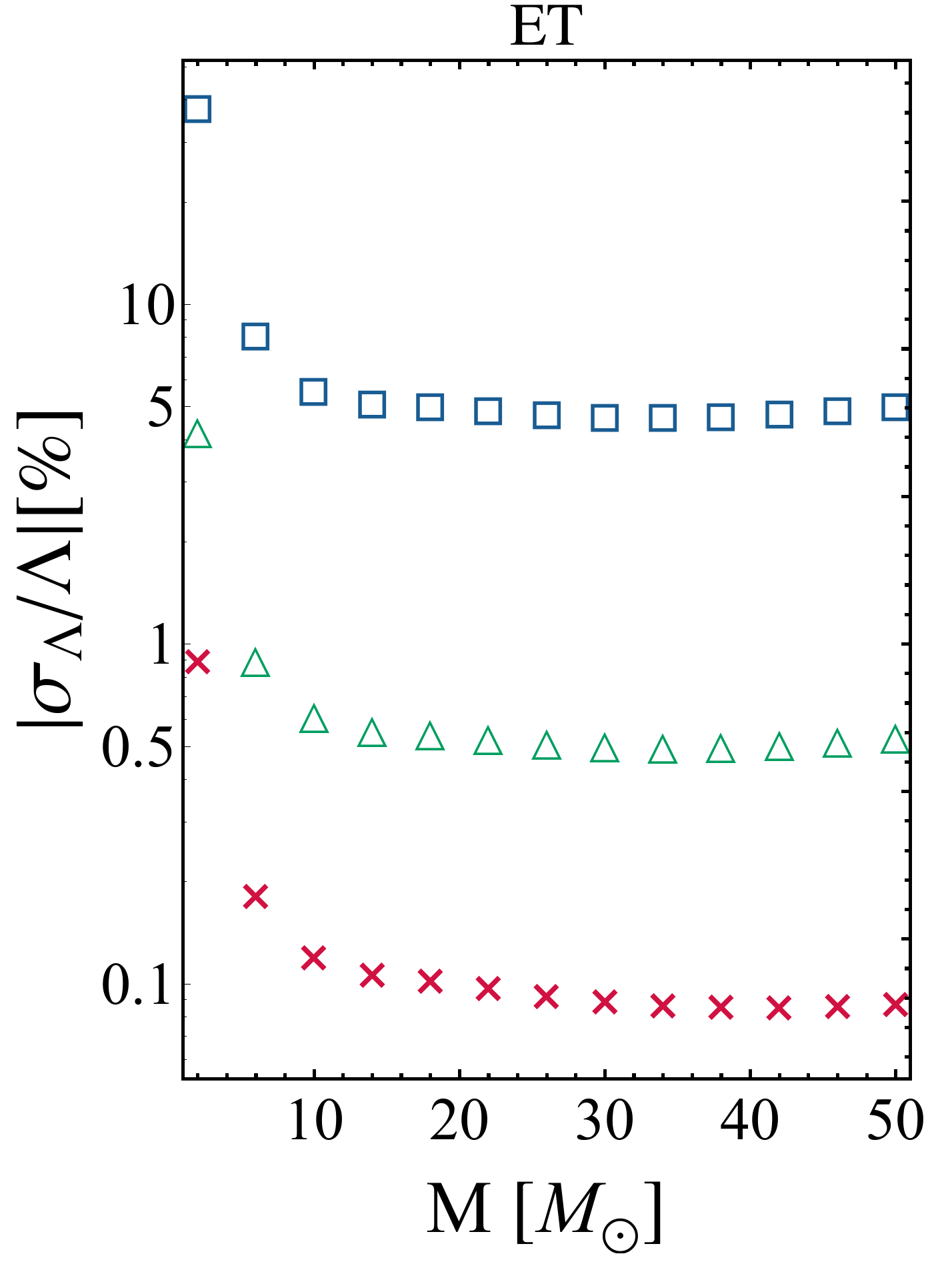}
\includegraphics[width=0.2625\textwidth]{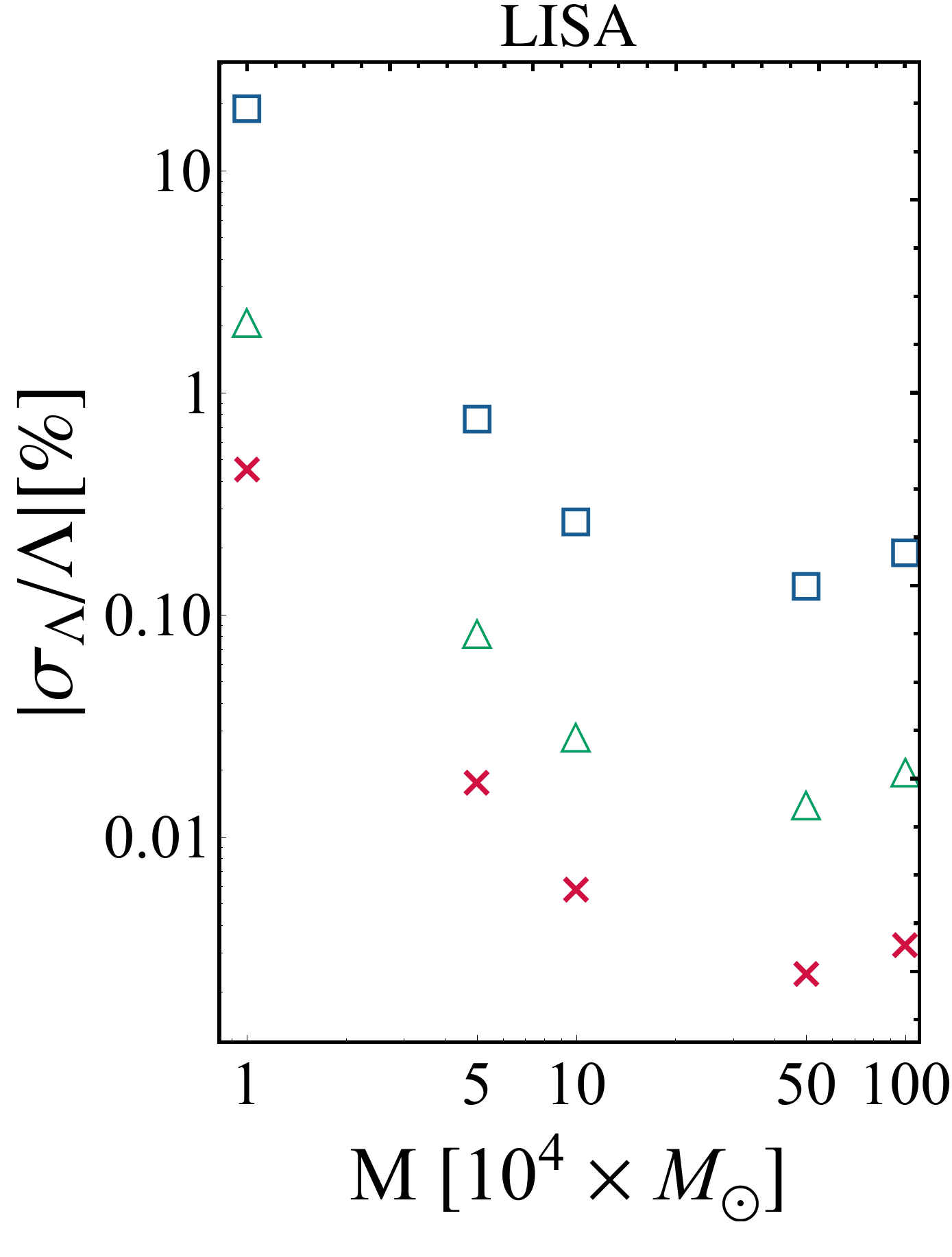}
\caption{Relative percentage errors on the average tidal deformability $\Lambda$ for BS-BS binaries observed by AdLIGO (left panel), 
ET (middle panel), and LISA (right panel), as a function of the BS mass and for different BS models considered in this work (for each model, we considered the most compact configuration in the stable branch; see main text for details). For terrestrial interferometers we assume a prototype binary at $d=100\,{\rm Mpc}$, while for LISA the source is located at $d=500\,{\rm Mpc}$.
The horizontal dashed line identifies the upper bound $\sigma_\Lambda/\Lambda=1$. Roughly speaking, a measurement of the TLNs for systems which lie below the threshold line would be incompatible with zero and, therefore, the corresponding BSs can be distinguished from BHs. Here $\Lambda$ is given by Eq.~\eqref{deflambda}, the two inspiralling objects have the same mass, and $\sigma_\Lambda/\Lambda\sim \sigma_{k_2^E}/{k_2^E}$.
}
\label{fig:detectabilityBS}
\end{figure*}

Our main results are summarized in Table~\ref{tab:summary} and in Fig.~\ref{fig:detectabilityBS}, and are discussed in detail in the rest of the paper.
\begin{table*}[ht!]
\centering
\caption{\footnotesize Tidal Love numbers (TLNs) of some exotic compact objects (ECOs) and BHs in Einstein-Maxwell theory and modified theories of gravity; details are given in the main text. 
As a comparison, we provide the order of magnitude of the TLNs for static NSs with compactness $C\approx0.2$ (the precise number depends on the neutron-star equation of state; see Table~\ref{tab:EOS} for more precise fits).
For BSs, the table provides the lowest value of the corresponding TLNs among different models (cf.~Sec.~\ref{sec:BSs}) and values of the compactness. In the polar case, the lowest TLNs correspond to solitonic BSs with compactness $C\approx 0.18$ or $C\approx0.20$ (when the radius is defined as that containing 99\% or 90\% of the total mass, respectively). In the axial case, the lowest TLNs correspond to a massive BS with $C\approx 0.16$ or $C\approx0.2$ (again for the two definitions of the radius, respectively) and in the limit of large quartic coupling.
%
For other ECOs, we provide expressions for very compact configurations where the surface $r_0$
sits at $r_0\sim 2M$ and is parametrized by $\xi:=r_0/(2M)-1$; the full results are available online~\cite{webpage}. 
In the Chern-Simons case, the axial $l=3$ TLN is affected by some ambiguity and is denoted by a question mark [see Sec.~\ref{sec:Chern} for more details]. Note that the TLNs for Einstein-Maxwell and Chern-Simons gravity were obtained under the assumption of vanishing electromagnetic and scalar tides.
}
\label{tab:summary}
\begin{tabular}{|c|c|cccc|}
\hline
\hline
                                       &                  & \multicolumn{4}{c|}{Tidal Love numbers} \\
                                       &                  & $k_2^E$      &   $k_3^E$   & $k_2^B$      & $k_3^B$     \\
\hline
\multirow{1}{*}{NSs}                   & 	          &  $210$        	 &   $1300$      &   $11$       &       $70$  \\
                                       \hline
\multirow{4}{*}{ECOs}                  & Boson star        & $41.4$    	 & $402.8$  & $-13.6$  & $-211.8$ \\
                                       & Wormhole          & $\frac{4}{5(8+3\log\xi)}$         	 & $\frac{8}{105(7+2\log\xi)}$         & $\frac{16}{5(31+12\log\xi)}$          & $\frac{16}{7(209+60\log\xi)}$         \\
                                       & Perfect mirror    & $\frac{8}{5 (7+3 \log\xi)}$         	 & $\frac{8}{35 (10+3 \log\xi)}$          & $\frac{32}{5 (25+12 \log\xi)}$          & $\frac{32}{7 (197+60 \log\xi)}$ \\
                                       & Gravastar         &  $\frac{16}{5 (23-6\log{2}+9 \log\xi)}$         	 &  $\frac{16}{35 (31-6 \log{2}+9 \log\xi)}$       &  $\frac{32}{5 (43-12\log{2}+18\log\xi)}$        &   $\frac{32}{7 (307-60 \log{2}+90 \log\xi)}$      \\
                                       \hline
\multirow{4}{*}{BHs} & Einstein-Maxwell &  $0$      	 & $0$     & $0$     & $0$     \\
                                       & Scalar-tensor    &  $0$      	 & $0$     & $0$     & $0$     \\
                                       & Chern-Simons     &  $0$      	 & $0$        &   $1.1\frac{\alpha^2_{\rm CS}}{M^4}$      &  $11.1\frac{\alpha^2_{\rm CS}}{M^4}$?       \\
\hline
\hline
\end{tabular}    
\end{table*}
Table~\ref{tab:summary} lists the lowest quadrupolar ($l=2$) and octupolar ($l=3$) polar and axial TLNs for various models of ECOs in GR,
and for some static BHs in other gravity theories.
Table~\ref{tab:summary} also compares the TLNs of these objects with the corresponding ones for a typical NS (cf.\ also Table~\ref{tab:EOS} in Appendix~\ref{section:TLN_stars}).

One of our main results is that the TLNs of several ECOs display a logarithmic dependence in the BH limit, i.e. when the compactness of the object approaches that of a BH,
\be
C:=M/r_0\to 1/2\,,
\ee
where $M$ and $r_0$ are the mass and the radius of the object. As shown in Table~\ref{tab:summary}, this property holds for wormholes, thin-shell gravastars, and for a simple toy model of a static object with a perfectly reflecting surface~\cite{Saravani:2012is,Abedi:2016hgu}.
It is natural to conjecture that this logarithmic behavior is model independent and will hold for any ECO whose exterior spacetime is arbitrarily close to that of a BH in the $r_0\to 2M$ limit. This mild dependence implies that even the TLNs of an object with $r_0-2M\approx \ell_P$ are not extremely small, contrarily to what one could expect. Indeed, we estimate that the dimensionless TLNs defined in Eq.~\eqref{Lovenumbersdef1} below are 
\begin{equation}
 k_2^{E,B}\approx {\cal O}(10^{-3})\,, \qquad k_3^{E,B}\approx {\cal O}(10^{-4})\,, \label{estimates}
\end{equation}
for an ECO with $r_0-2M\approx \ell_P$ and in the entire mass range $M\in [1,100]\, M_\odot$. Note that, with the exception of polar TLNs of boson stars, all TLNs of ultracompact exotic objects listed in Table~\ref{tab:summary} have the opposite sign relative to the neutron-star case (cf.\ the discussion in Refs.~\cite{Pani:2015tga,Uchikata:2016qku} applied to a particular model). Negative values of TLNs were also found previously for ultracompact anisotropic NSs~\cite{Yagi:2016ejg}.   

Furthermore, we show that the TLNs of a charged BH in Einstein-Maxwell theory and of an uncharged static BH in Brans-Dicke theory vanish, as in GR, whereas the TLNs of a BH in Chern-Simons gravity are nonzero, even though the static BH solution to this theory is described by the Schwarzschild metric. The results for Einstein-Maxwell and Chern-Simons gravity were obtained with the assumption that there are no electromagnetic and scalar tidal fields. As expected, the TLNs are proportional to the coupling constant of the theory so that any constraint on them can be potentially converted into a test of gravity.

The accuracy with which GW detectors can estimate the TLNs of compact objects is shown in Fig.~\ref{fig:detectabilityBS} for three BS models, where the two inspiralling objects are assumed to be equal. 
In moderately optimistic scenarios, a GW detection of a compact-binary coalescence with LIGO can place an upper bound on the TLNs of the two objects at the level of $k_2^{E}\sim 10$, whereas the future Einstein Telescope~(ET)~\cite{Punturo:2010zz} can potentially improve this constraint by almost a factor of a hundred. Interestingly, the future space interferometer LISA~\cite{AmaroSeoane:2012je} has the ability to set much tighter constraints [cf. also Fig.~\ref{fig:detectabilitygeneric}] and to rule out several candidates of supermassive ECOs.
In essence, both Earth and space-based detectors are able to discriminate even the most
compact BSs, by imposing stringent bounds on their TLNs. By contrast, as we show in Sec.~\ref{sec:detectability}, only LISA is able to probe the regime of very compact ECOs, describing geometries which are microscopic corrections at the horizon scale, for which the compactness $C=0.48$ or higher.

\section{Setup: Tidal Love numbers of static objects} \label{setup}
Let us consider a compact object immersed in a tidal environment~\cite{PoissonWill}. Following Ref.~\cite{Binnington:2009bb}, we define the symmetric and trace-free
polar and axial\footnote{It is slightly more common to use the distinction electric/magnetic components rather than polar/axial. Since we shall discuss also electromagnetic fields, we prefer to use the former distinction.} tidal multipole moments of order $l$ as ${\cal E}_{a_1\dots a_l}\equiv [(l-2)!]^{-1}\langle
C_{0a_10a_2;a_3\dots a_l}\rangle$ and
${\cal B}_{a_1\dots a_l}\equiv [\frac{2}{3}(l+1)(l-2)!]^{-1}\langle\epsilon_{a_1 b c} C^{bc}_{a_2 0;a_3\dots a_l}\rangle$,
where $C_{abcd}$ is the Weyl tensor, a semicolon denotes a covariant derivative, $\epsilon_{abc}$ is the permutation
symbol, the angular brackets denote symmetrization of the indices $a_i$ and all traces are removed. 
The polar (respectively, axial) moments ${\cal E}_{a_1\dots a_l}$ (respectively, ${\cal B}_{a_1\dots a_l}$) can be decomposed in a basis of even (respectively, odd) parity spherical harmonics. 
We denote by ${\cal E}^{lm}$ and  ${\cal B}^{lm}$ the amplitudes of the polar and axial components of the external tidal field with harmonic indices $(l,m)$, where $m$ is the azimuthal number ($|m|\leq l$). The structure of the external tidal field is entirely encoded in the coefficients ${\cal E}^{lm}$ and  ${\cal B}^{lm}$ (cf. Ref.~\cite{Binnington:2009bb} for details).

As a result of the external perturbation, the mass and current multipole moments\footnote{We adopt the Geroch-Hansen definition of multipole moments~\cite{Geroch:1970cd,Hansen:1974zz}, equivalent~\cite{Gursel:1983} to the one by Thorne~\cite{Thorne:1980ru} in asymptotically mass-centered Cartesian coordinates.} ($M_l$ and $S_l$, respectively) of the compact object will be deformed. In linear perturbation theory, these deformations are proportional to the applied tidal field. In the nonrotating case, mass (current) multipoles have even (odd) parity, and therefore they 
only depend on polar (axial) components of the tidal field.\footnote{This symmetry is broken if the compact object is spinning due to spin-tidal couplings. In such case, there exists a series of selection rules that allow to define a wider class of ``rotational'' TLNs~\cite{Pani:2015hfa,Landry:2015zfa,Pani:2015nua,Landry:2015snx}. In this paper, we neglect spin effects to leading order.} Hence, we can define the (polar and axial) TLNs as~\cite{Hinderer:2007mb,Binnington:2009bb}
\begin{equation}
\begin{split}
&k^E_{l}\equiv -\frac12 \frac{l(l-1)}{M^{2l+1}} \sqrt{\frac{4\pi}{2l+1}}\frac{M_l}{\mathcal{E}_{l0}},\\
& k^B_{l}\equiv -\frac32 \frac{l(l-1)}{(l+1)M^{2l+1}}\sqrt{\frac{4\pi}{2l+1}}  \frac{S_l}{\mathcal{B}_{l0}}\,, \label{Lovenumbersdef1}
\end{split}
\end{equation}
where $M$ is the mass of the object, whereas $\mathcal{E}_{l0}$ (respectively, $\mathcal{B}_{l0}$) is the amplitude of the axisymmetric\footnote{We consider only nonspinning objects, hence the spacetime is spherically symmetric and, without loss of generality, we can define the TLNs in the axisymmetric ($m=0$) case. Clearly, this property does not hold when the object is spinning~\cite{Poisson:2014gka,Landry:2015zfa,Pani:2015nua}.} component of the polar (respectively, axial) tidal field.
The factor $M^{2l+1}$ was introduced to make the above quantities dimensionless. It is customary to normalize the TLNs by powers of the object's radius $R$ rather than by powers of its mass $M$. Here we adopted the latter nonstandard choice, since the radius of some ECOs (e.g.\, BSs) is not a well-defined quantity. Thus, our definition is related to those used by Hinderer, Binnington and Poisson (HBP)~\cite{Hinderer:2007mb,Binnington:2009bb} through
\be
\begin{split}
&k^{E,B}_{l\, {\rm ours}}=\left(\frac{R}{M}\right)^{2l+1} k^{E,B}_{l\, {\rm HBP}}\,. \label{conversion}
\end{split}
\ee

Modified theories of gravity and ECOs typically require the presence of extra fields which are (non)minimally coupled to the metric tensor. Here we shall consider some representative example of both scalar and vector fields.
A full treatment of this problem would require allowance for an extra degree of freedom, the external scalar and electromagnetic (EM) applied fields. It is generically expected that, in astrophysical situations, the ratio of a putative external (scalar or vector) field to the ordinary gravitational tidal field should be small. We will, therefore, focus only on situations where the only surviving field at large distances is gravitational.

We expand the metric, the scalar field, and the Maxwell field in spherical harmonics as presented in Appendix~\ref{app:comp}. Since the background is spherically symmetric, perturbations with different parity and different harmonic index $l$ decouple. In the following we discuss the polar and axial sector separately; due to the spherical symmetry of the background, the azimuthal
number $m$ is degenerate and we drop it.

Finally, in order to extract the tidal field and the induced multipole moments from the solution, we have adopted two (related) techniques. The first one relies on an expansion of the metric at large distances [cf.~Eqs.~\eqref{eq:gttexpansion} and~\eqref{eq:gtphiexpansion}] in terms of the multipole moments. The second technique relies on the evaluation of the Riemann tensor in Schwarzschild coordinates, whose tidal correction is related to the total tidal field in the local asymptotic rest frame~\cite{Fang:2005qq}. 
These two procedures agree with each other and --~at least in the case of ECOs~-- the computation of the TLNs is equivalent to the case of NSs~\cite{Hinderer:2007mb,Binnington:2009bb}. On the other hand, computing the TLNs of BHs in extensions of GR presents some subtleties which are discussed in Sec.~\ref{sec:BH}.

\section{Tidal perturbations of exotic compact objects} \label{sec:ECOs}
In this section we describe some representative models of ECOs and discuss their TLNs. Technical details are given in the Appendices.

\subsection{Boson stars} \label{sec:BSs}

\begin{table}[th]
\scriptsize
\begin{tabular}{|c|c|c|}
\hline\hline
Model & \begin{tabular}{@{}c@{}}Potential \\ $V(|\Phi|^2)$\end{tabular}  & \begin{tabular}{@{}c@{}}Maximum mass \\ $M_{\max}/M_{\odot}$\end{tabular}\\
\hline
Minimal & $\mu^2|\Phi|^2$ & $8\left(\frac{10^{-11}{\rm eV}}{m_S}\right)$\\
Massive & $\mu^2|\Phi|^2 + \frac{\alpha}{4}|\Phi|^4$ & $5\,\sqrt{\alpha\hbar}\left(\frac{0.1\,{\rm GeV}}{m_S}\right)^2$\\
Solitonic & $\mu^2|\Phi|^2\left[1-\frac{2|\Phi|^2}{\sigma_0^2}\right]^2$ & $5\left[\frac{10^{-12}}{\sigma_0}\right]^2\left(\frac{500\,{\rm GeV}}{m_S}\right)$\\
\hline\hline
\end{tabular}
\caption{Scalar potential and maximum mass for the BS models considered in this work. In our units, the scalar field $\Phi$ is dimensionless and the potential $V$ has dimensions of an inverse length squared. The bare mass of the scalar field is $m_S:=\mu\hbar$.
For minimal BSs, the scaling of the maximum mass is exact.
For massive BSs and solitonic BSs, the scaling of the maximum mass is approximate and holds only when $\alpha\gg \mu^2$ and when $\sigma_0\ll1$, respectively.}
\label{tab:BSs}
\end{table}

BSs are complex\footnote{If the scalar field is real, action~\eqref{actionKG} admits compact, self-gravitating, oscillating solutions known as oscillatons~\cite{Seidel:1991zh}. These solutions are metastable, but their decay time scale can largely exceed the age of the universe and their properties are very similar to those of BSs. We expect that the TLNs of BSs computed here are similar to those of an oscillaton star, although a detailed computation is left for future work.} bosonic configurations held together by gravity. In the simplest model they are solutions to the Einstein-Klein-Gordon theory,
\begin{equation}
S=\int d^4 x \sqrt{-g} \left[\frac{R}{16\pi} -g^{ab}\partial_a\Phi^*\partial_b\Phi-V\left(|\Phi|^2\right)\right] \,.\label{actionKG}
\end{equation}
%
BSs have been extensively studied in the past and have been proposed as BH mimickers and dark matter candidates,
see e.g.\ Refs.~\cite{Jetzer:1991jr,Schunck:2003kk,Liebling:2012fv,Macedo:2013qea,Brito:2015yfh}.

BSs are typically classified according to the scalar potential in the above action; here we investigate three of the most common models:
minimal BSs~\cite{Kaup:1968zz,Ruffini:1969qy}, massive BSs~\cite{Colpi:1986ye} and
solitonic BSs~\cite{Friedberg:1986tq}. The corresponding scalar potential for these models and the maximum mass for nonspinning solutions are listed in Table~\ref{tab:BSs}. A more comprehensive list of BS models can be found in Ref.~\cite{Liebling:2012fv}.

Depending on the model, compact BSs with masses comparable to those of ordinary stars or BHs require a certain range of the scalar mass $m_S:= \mu \hbar$.
For minimal BSs, the maximum mass in Table~\ref{tab:BSs} is comparable to the Chandrasekhar limit for NSs only for an ultralight field with $m_S\lesssim 10^{-11}\,{\rm eV}$.
For massive BSs, the maximum mass is of the same order of the Chandrasekhar
limit if $m_S\sim 0.1\,{\rm GeV}$ and the quartic coupling is large, $\alpha\hbar\sim1$~\cite{Colpi:1986ye}.
Finally, solitonic BSs may reach massive ($M\gtrsim M_{\odot}$) or supermassive ($M\gtrsim 10^6 M_{\odot}$) configurations even for
heavy bosons with $m_S\sim 500\,{\rm GeV}$ if the coupling parameter in their potential is $\sigma_0\lesssim 10^{-12}$ or $\sigma_0\lesssim 10^{-15}$, respectively~\cite{Friedberg:1986tq}. For massive and solitonic BSs, the scaling of the maximum mass in Table~\ref{tab:BSs} is approximate and valid only when $\alpha\gg \mu^2$ and when $\sigma_0\ll1$, respectively. In our numerical analysis, we have considered $\alpha=10^4\mu^2$ and $\sigma_0=0.05$, whereas the mass term $\mu$ can be rescaled away (cf., e.g., discussion in Ref.~\cite{Macedo:2013jja}).

Even though BSs have a wide range of compactness, which depends basically on their total mass (cf.~ Fig.~\ref{fig:massradius} in Appendix~\ref{app:BS}),
interactions between BSs typically leads to a net weight gain, clustering old BSs close to the mass peak~\cite{Brito:2015yfh}, which also coincides with the peak of compactness.

The details of the numerical procedure to compute the TLNs of a BS are presented in Appendix~\ref{app:BS}.
Figure~\ref{fig:kBS} shows the TLNs of the BS models presented above as a function of the total mass $M$, the latter being normalized by the total mass $M_{\rm max}$ of the corresponding model. We only show static configurations in the stable branch, i.e. with a mass smaller than $M_{\rm max}$ (cf.\ discussion in Appendix~\ref{app:BS}).
\begin{figure*}[ht]
\includegraphics[width=0.45\textwidth]{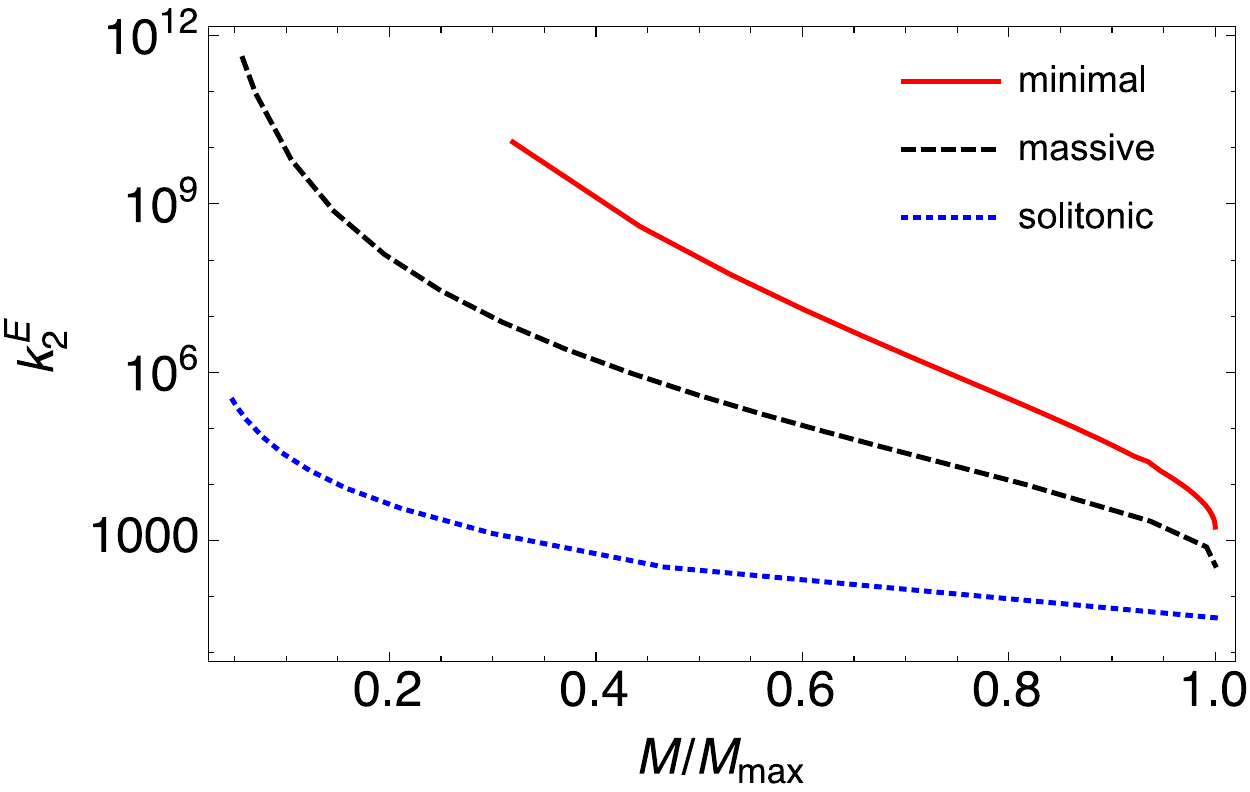}\qquad
\includegraphics[width=0.45\textwidth]{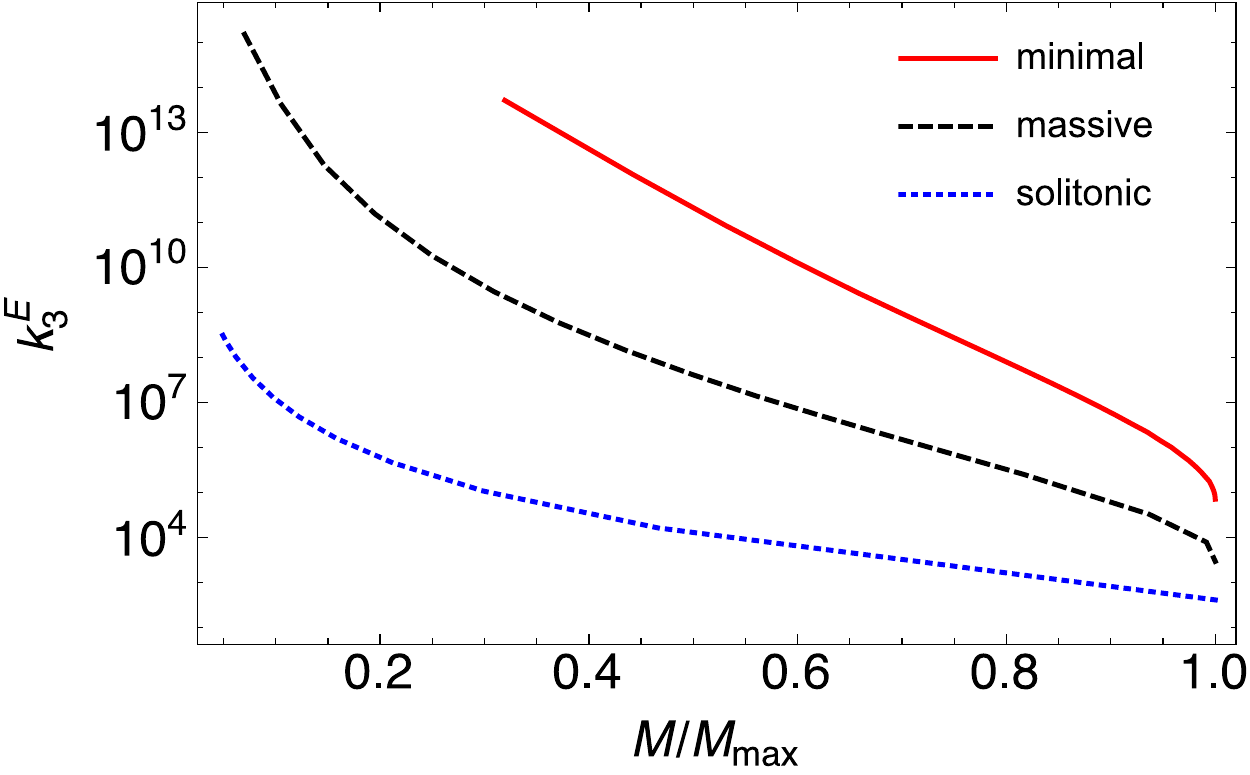}\\
\includegraphics[width=0.45\textwidth]{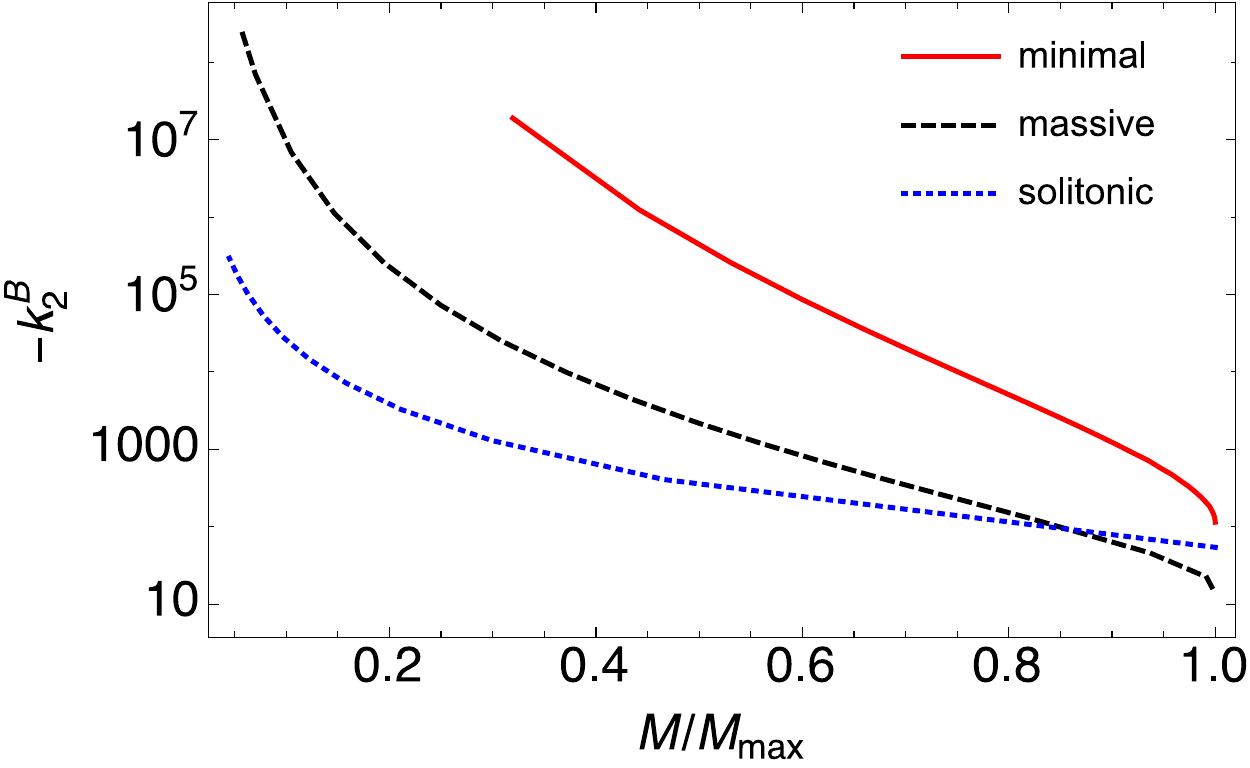}\qquad
\includegraphics[width=0.45\textwidth]{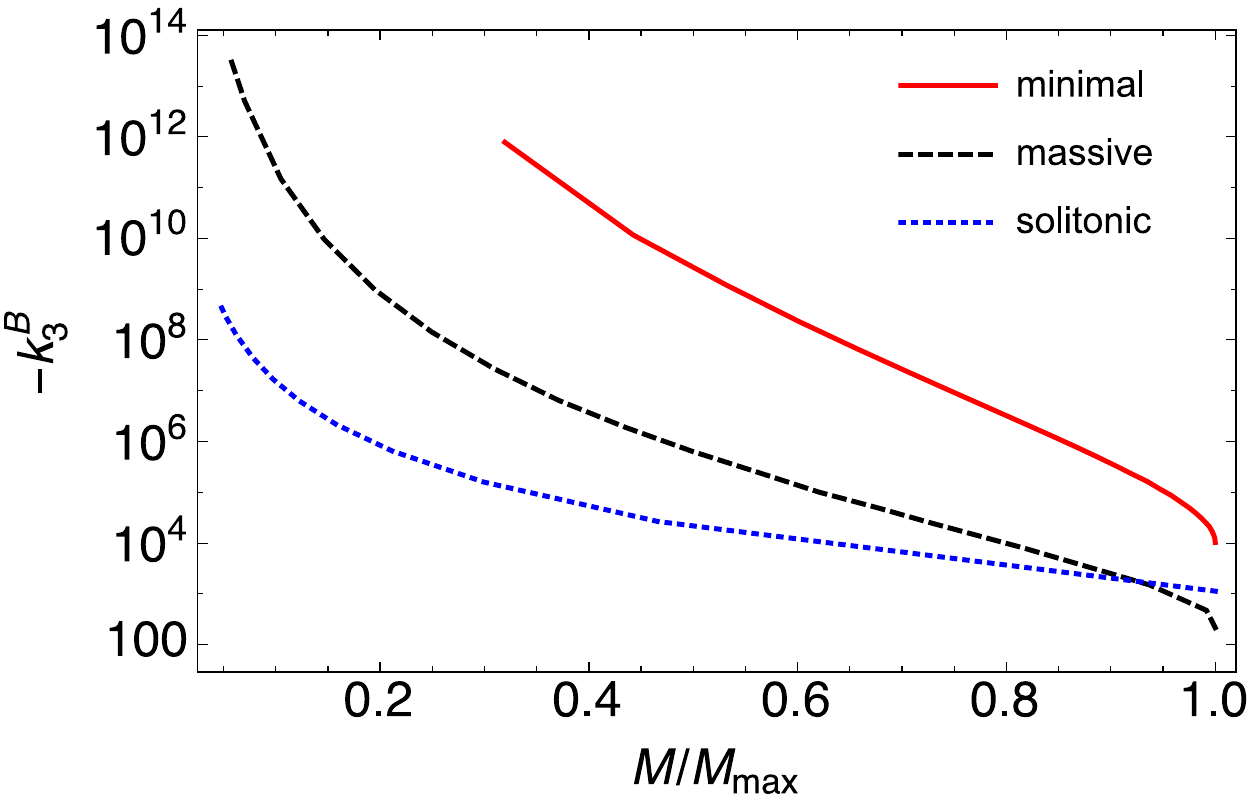}
\caption{Polar (top panels) and axial (bottom panels) TLNs for minimal, massive and solitonic BSs. Left and right panels refers to $l=2$ and $l=3$, respectively. For massive and solitonic BSs we have considered $\alpha=10^4\mu^2$ and $\sigma_0=0.05$, respectively. With these values, the maximum mass scales approximately as shown in Table~\ref{tab:BSs}. Numerical data are available online~\cite{webpage}. These plots include only stars in the stable branch.}\label{fig:kBS}
\end{figure*}
For minimal BSs and for $l=2$ polar case, our results agree with those recently obtained in Ref.~\cite{Mendes:2016vdr}.
In addition, we also present the results for $l=2$ and $l=3$, for both axial and polar TLNs, and for the three BS models previously discussed. 

The behavior of the TLNs of BSs is in qualitative agreement with that of NSs. For a given BS model with a given mass, the magnitude of the polar TLN is larger than that of an axial TLN with the same~$l$. Furthermore, in the Newtonian regime ($M\to0$) the TLNs scale as $k_l^E\sim C^{-(2l+1)}$ and $k_l^B\sim -C^{-2l}$. This scaling is in agreement with the neutron-star case (cf.\ Ref.~\cite{Binnington:2009bb} and Table~\ref{tab:EOS}), whereas the sign of the axial TLNs is opposite. Finally, all TLNs are monotonic functions of the compactness, so that more compact configurations have smaller tidal deformability. The phenomenological implications of these results are discussed in Sec.~\ref{sec:detectability}.

\subsection{Models of microscopic corrections at the horizon scale} \label{sec:QC}
Several phenomenological models of quantum BHs introduce a Planck-scale modification near the horizon. In this section, we consider three toy models for microscopic corrections at the horizon scale, namely a wormhole~\cite{visser1995lorentzian}, a Schwarzschild geometry with a perfectly reflective surface near the horizon~\cite{Saravani:2012is,Abedi:2016hgu}, and a thin-shell gravastar~\cite{Mazur:2001fv}.\footnote{Many of these objects are unstable or require exotic matter distributions. We will not be concerned with these issues here.} These models have some common features: (i)~the exterior spacetime is described by the Schwarzschild metric; (ii)~the interior is either vacuum or de Sitter and the tidal perturbation equations can be solved for in closed form; (iii)~simple junction or boundary conditions at the radius $r_0$ of the object can be imposed to connect the perturbations in the interior with those in the exterior. As a result of these properties, the TLNs of these models can be computed in closed analytical form. As we show, the qualitative features are the same and --~especially in the BH limit~-- do not depend strongly on the details of the models. Below, we present explicit formulas for the BH limit, expressions for generic compactness are provided online~\cite{webpage}. The details of the computation are given in Appendix~\ref{app:ECOs}.

\subsubsection{Wormholes}\label{sec:WHs}

The simplest models of wormhole solutions consist in taking two copies of the ordinary Schwarzschild solution and remove from them the four-dimensional regions described by $r_{1,2}\leq r_0$~\cite{visser1995lorentzian}.
With this procedure, we obtain two manifolds whose geodesics terminate at the timelike hypersurfaces
\begin{equation}
\partial\Omega_{1,2}\equiv\left\lbrace r_{1,2}= r_0\, |\, r_0>2M\right\rbrace\,.
\end{equation}
The two copies are now glued together by identifying these two boundaries, $\partial\Omega_1=\partial\Omega_2$, such that the resulting spacetime is geodesically complete and comprises of two distinct regions connected by a wormhole with a throat at $r=r_0$. Since the wormhole spacetime is composed by two Schwarzschild metrics, the stress-energy tensor vanishes everywhere except on the throat of the wormhole. The patching at the throat requires a thin shell of matter with surface density and surface pressure
\begin{equation}
\sigma=-\frac{1}{2\pi r_0} \sqrt{1-\frac{2M}{r_0}}\,,\quad p=\frac{1}{4\pi r_0}\frac{1-M/r_0}{\sqrt{1-2M/r_0}}\,,
\end{equation}
which imply that the weak and the dominant energy conditions are violated, whereas the null and the strong energy conditions are satisfied when $r_0<3M$~\cite{Cardoso:2016oxy}.
To cover the two patches of the spacetime, we use the radial tortoise coordinate $r_*$, which is defined by
\begin{equation}
\label{tortoise}
\frac{dr}{dr_*}=\pm\left(1-\frac{2M}{r}\right)\,,
\end{equation} 
where the upper and lower sign refer to the two sides of the wormhole. Without loss of generality, we can assume that the tortoise coordinate at the throat is zero, \mbox{$r_*(r_0)=0$}, so that one side corresponds to $r_*>0$ whereas the other side corresponds to $r_*<0$.

\begin{figure}[th]
\centering
\includegraphics[width=0.45\textwidth]{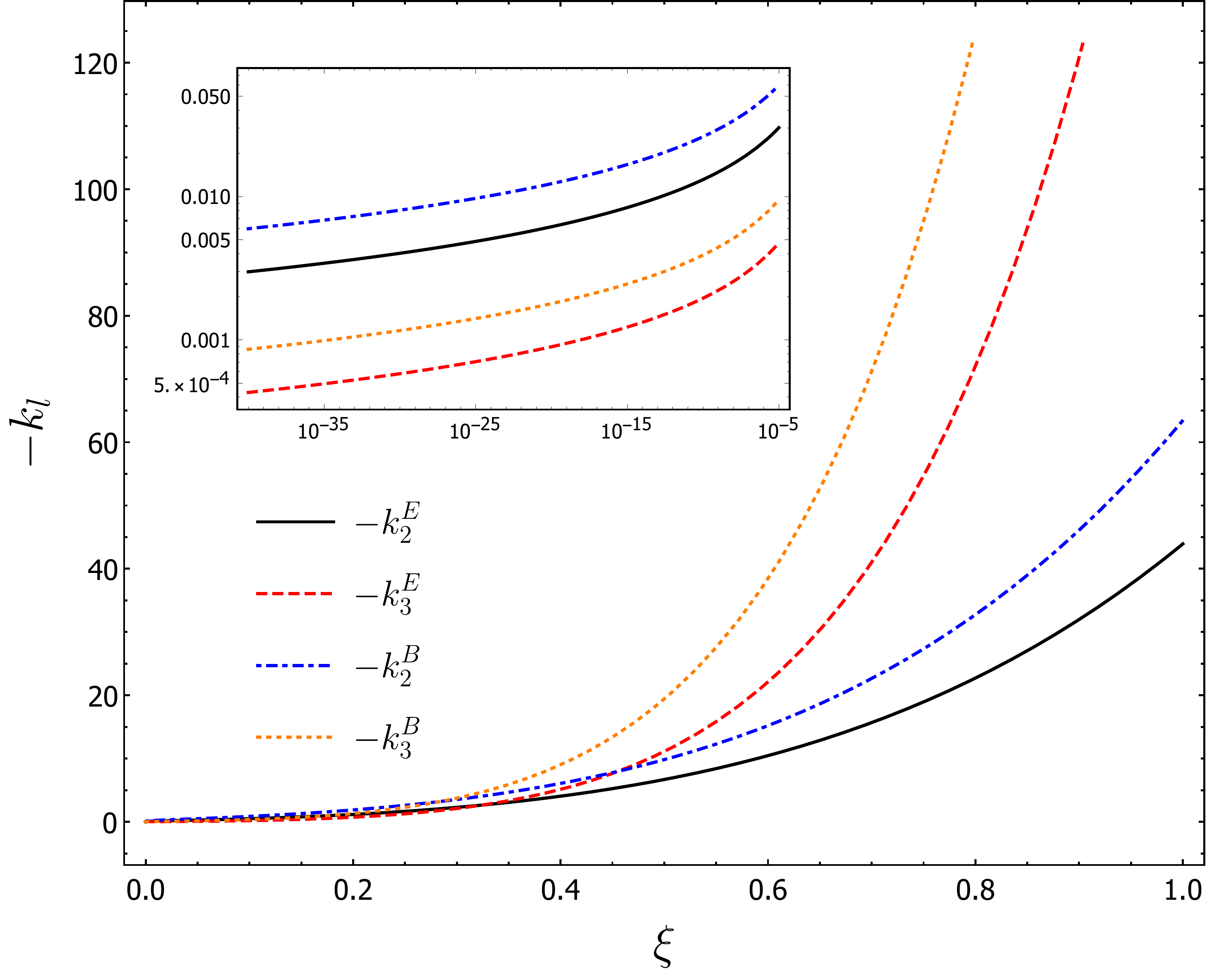}
\caption{The $l=2$ and $l=3$, axial- and polar-type TLNs for a stiff wormhole constructed by patching two Schwarzschild spacetimes at the throat radius $r=r_0>2M$. The TLNs are negative and all vanish in the BH limit, $r_0\to 2M$. The latter is better displayed in the inset.}
\label{Lovewormhole}
\end{figure}

In Fig.~\ref{Lovewormhole}, we show the polar and axial TLNs with \mbox{$l=2,3$} as functions of $\xi:=r_0/(2M)-1$. Interestingly, in this case the TLNs have the opposite sign to those of a NS. Furthermore, they vanish in the BH limit, i.e. when $r_0\to 2M$ or $\xi\to0$. The behavior of the TLNs in the BH limit reads
\beq
k^E_2&\sim& \frac{4}{5(8+3\log\xi)}\,,\\
k^E_3&\sim& \frac{8}{105(7+2\log\xi)}\,,\\
k^B_{2}&\sim&\frac{16}{5(31+12\log\xi)}\,,\\
k^B_{3}&\sim&\frac{16}{7(209+60\log\xi)}\,,
\eeq
where we have omitted subleading terms of ${\cal O}\left(\frac{\xi}{(\log\xi)^2}\right)$. On the other hand, in the Newtonian limit we get $k^{E,B}_l\sim C^{-(2l+1)}$. Interestingly, while the scaling for polar TLNs agrees with that of NSs (cf.\ Table~\ref{tab:EOS}), that for the axial TLNs is different.

The logarithmic dependence of the TLNs is very interesting, because it implies that the deviations from zero (i.e., from the BH case) are relatively large even when the throat is located just a Planck length away from the would-be horizon $r_0-2M\sim \ell_P\approx 1.6\times 10^{-33}\,{\rm cm}$. In this case, the above results yield 
\begin{equation}
\begin{split}
 k_2^{E}&\approx -3\times 10^{-3}\,,\qquad  k_2^{B}\approx -6\times 10^{-3}\,, \\
 k_3^{E}&\approx -4\times 10^{-4}\,, \qquad k_3^{E}\approx -9\times 10^{-4}\ ,
 \end{split}
\end{equation}
for a wormhole in the entire mass range $M\in [1,100]\, M_\odot$. 

\subsubsection{Perfectly-reflective mirror} \label{sec:Z2}
Thermodynamical arguments suggest that any horizonless microscopic model of BH should act as a mirror, at least for
long wavelength perturbations~\cite{Saravani:2012is,Abedi:2016hgu}.
Motivated by this scenario, we consider a Schwarzschild geometry with a perfect mirror at $r=r_0>2M$ and impose Dirichlet boundary conditions on the Regge-Wheeler and Zerilli functions, for the axial and polar sector, respectively. Thus, our strategy is to consider the stationary limit of generically dynamical perturbations (in the Fourier space, where $\omega$ is the frequency of the perturbation) of a Schwarzschild geometry.

\begin{figure}[th]
\includegraphics[width=0.45\textwidth]{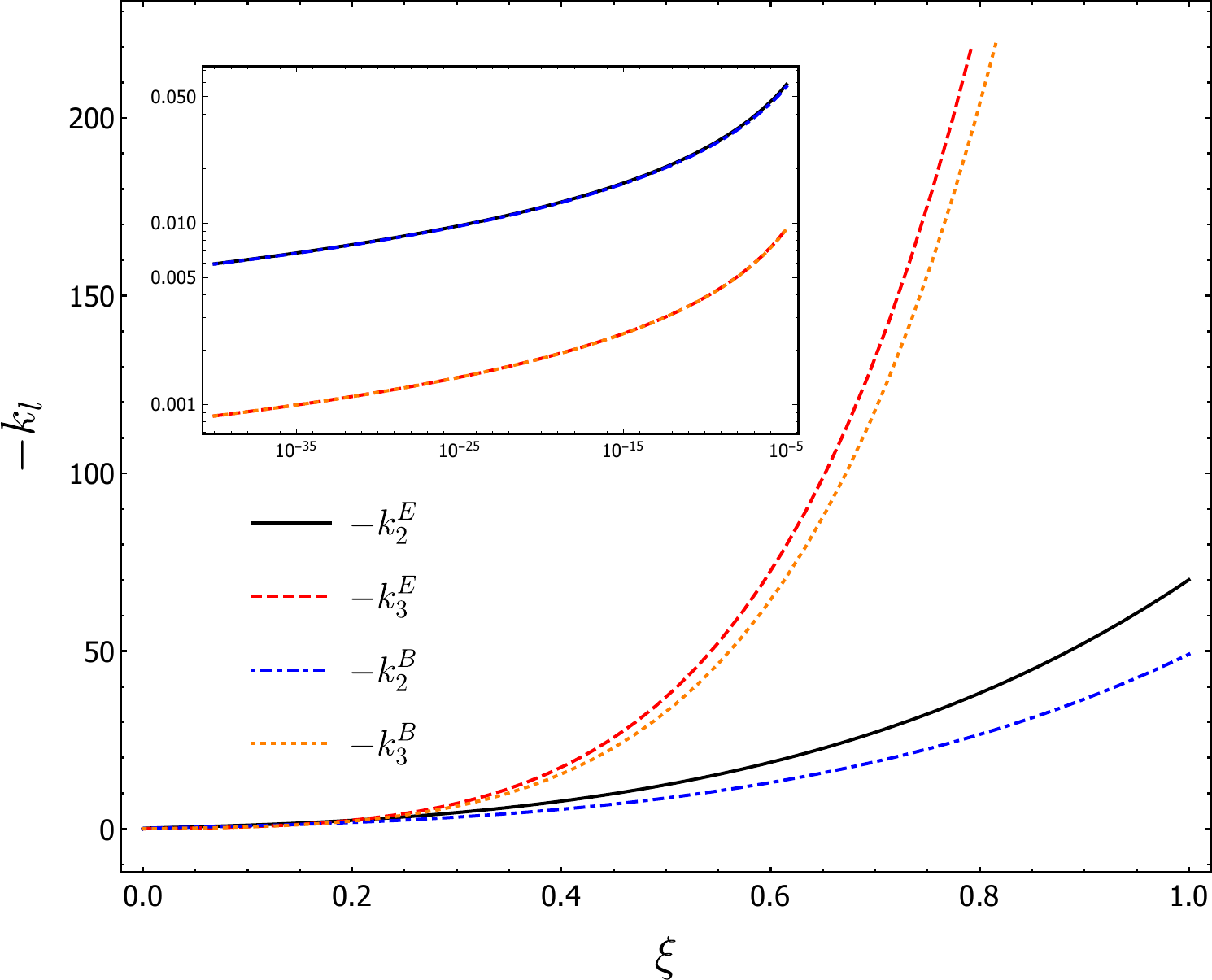}
\caption{TLNs for a toy model of Schwarzschild metric with a perfectly reflective surface at $r=r_0>2M$. The TLNs are all negative and vanish in the BH limit, $r_0\to 2M$. Close to the BH limit, the polar- and axial-type Love numbers for the same multipolar order are almost identical, as shown in the inset.}\label{fig:mirror}
\end{figure}

The final result, in the $\xi\to0$ limit, reads (cf.\ Appendix~\ref{app:ECOs} for details)
\begin{eqnarray}
 k_2^E &\sim& \frac{8}{5 (7+3 \log\xi)} \approx -6\times 10^{-3}\,,\\
 k_3^E &\sim& \frac{8}{35 (10+3 \log\xi)}\approx -9\times 10^{-4}\,, \\
 k_2^B &\sim&\frac{32}{5 (25+12 \log\xi)}\approx-6\times10^{-3}\,,\\
 k_3^B &\sim&\frac{32}{7 (197+60 \log\xi)}\approx -9\times 10^{-4}\,,
\end{eqnarray}
where the last step is evaluated at $r_0-2M\sim \ell_P\approx 1.6\times 10^{-33}\,{\rm cm}$ and it is roughly valid in the entire mass range $M\in [1,100]\, M_\odot$ due to the mild logarithmic dependence.
We note that also for this model all TLNs are negative (i.e., they have the opposite sign relative to the neutron-star case) and that $k^{E,B}_l\sim C^{-(2l+1)}$ in the Newtonian limit. The TLNs $k_l^{E,B}$ for this model as functions of the compactness are shown in Fig.~\ref{fig:mirror}.

\subsubsection{Thin-shell gravastars} \label{sec:GSs}
For completeness, here we briefly consider the case of another ECO, namely gravastars~\cite{Mazur:2001fv}. The interior of these objects is described by a patch of de Sitter space, which is smoothly connected to the Schwarzschild exterior through an intermediate region filled with a perfect fluid. A particularly simple model is the so-called thin-shell gravastar~\cite{Visser:2003ge}, in which the thickness of the intermediate region shrinks to zero. Remarkably, these models are simple enough that the TLNs can be computed analytically~\cite{Pani:2015tga,Uchikata:2016qku}.

Among thin-shell gravastars, we consider the simplest case where the background metric is  Eq.~\eqref{staticmetric} with,
\begin{equation}
e^{\Gamma}=e^{-\Lambda_g}=\left\{\begin{array}{l}
                                1-\frac{2M}{r} \quad \;\;\; r>r_0\\
                                1-2C\frac{r^2}{r_0^2} \quad r<r_0
                               \end{array}\right. \,.
\end{equation}
In this model, the thin shell is described by a fluid with zero energy density and negative pressure~\cite{Pani:2015tga}. 
By extending the formalism developed in Ref.~\cite{Uchikata:2016qku} (cf.\ also Refs.~\cite{Uchikata:2015yma,Pani:2015tga}), it is easy to compute the TLNs of this solution.
In the BH limit, the computation derived in Appendix~\ref{app:ECOs} yields\footnote{This result corrects the computation performed in Ref.~\cite{Pani:2015tga}, which is flawed due to the fact that it does not impose the correct boundary conditions across the shell. For a stiff equation of state, the correct boundary conditions read $[[K]]=0=[[d K/dr_*]]$ as derived in Appendix~\ref{app:ECOs} and in Ref.~\cite{Uchikata:2016qku}.}
\begin{eqnarray}
 k_2^E &\sim& \frac{16}{5 (23-6\log{2}+9 \log\xi)}\approx-4\times10^{-3}\,,\\
 k_3^E &\sim& \frac{16}{35 (31-6 \log{2}+9 \log\xi)} \approx -6\times 10^{-4}\,,\\
 k_2^B &\sim&\frac{32}{5 (43-12\log{2}+18\log\xi)}\approx-4\times10^{-3}\,,\\
 k_3^B &\sim&\frac{32}{7 (307-60 \log{2}+90 \log\xi)}\approx -6\times 10^{-4}\,.
\end{eqnarray}
As already noted in Ref.~\cite{Pani:2015tga}, the Newtonian regime of a gravastar is peculiar due to the de Sitter interior; consequently, the TLNs scale as $k_l^{E,B}\sim -C^{-2l}$. In this case, the scaling of the polar TLNs is different from that of an ordinary NS and of the other models of microscopic corrections at the horizon scale, whereas the scaling of the axial TLNs is the same as that for ordinary NSs and BSs.

\begin{figure}[ht]
\includegraphics[width=0.45\textwidth]{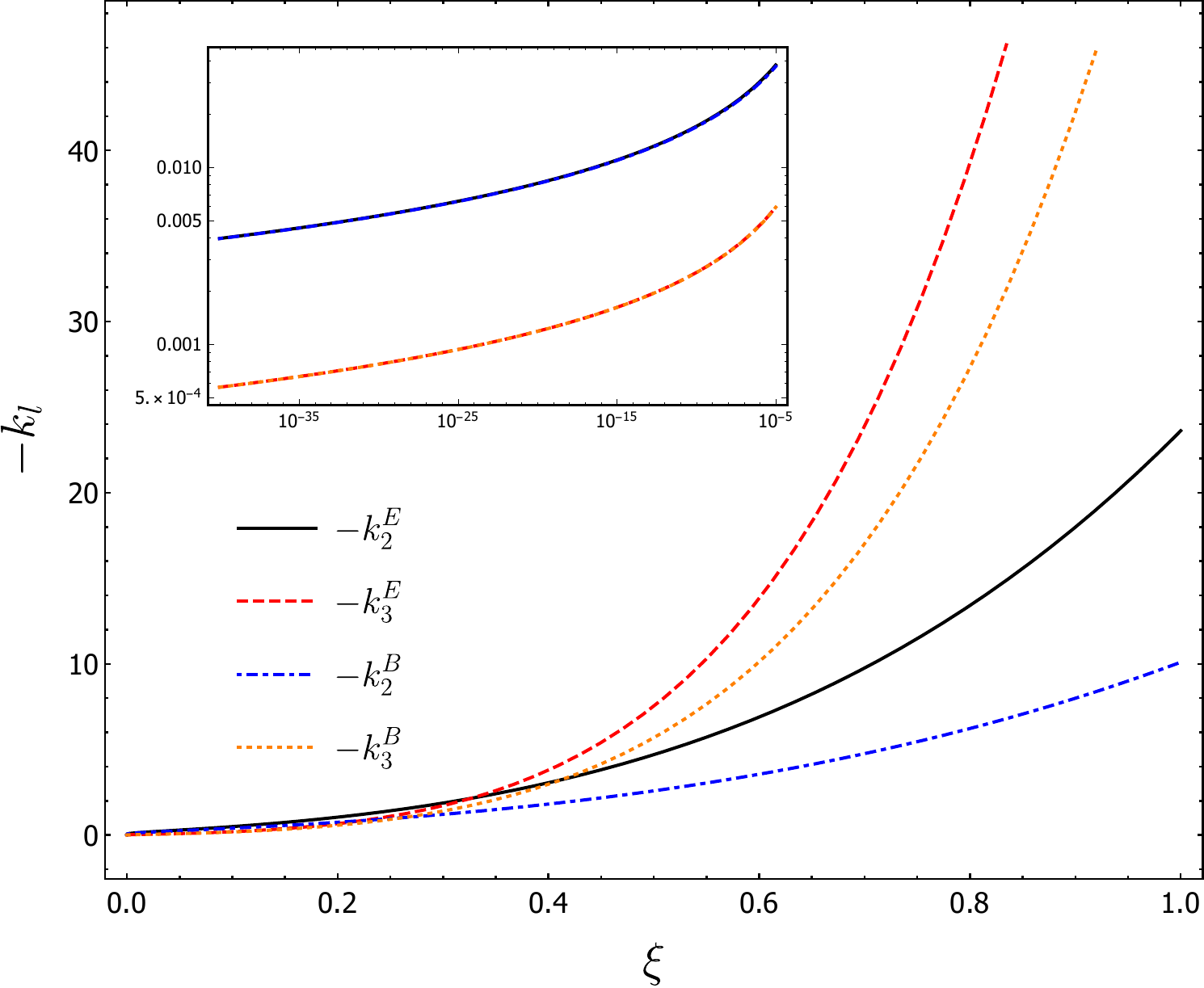}
\caption{TLNs for a thin-shell gravastar with zero energy density as a function of the compactness. More generic gravastar models are presented in Ref.~\cite{Uchikata:2016qku}. The TLNs are all negative and vanish in the BH limit, $r_0\to 2M$. Similar to the perfectly-reflective mirror case, the polar- and axial-type Love numbers for the same multipolar order coincide in the BH limit, as shown in the inset.}\label{fig:kGS}
\end{figure}

Interestingly, also in the gravastar case, the axial and polar TLNs have a logarithmic behavior in the BH limit and they are negative, as discussed in Refs.~\cite{Pani:2015tga,Uchikata:2016qku} for the polar case only. The behavior of $k_l^{E,B}$ as functions of the compactness is shown in Fig.~\ref{fig:kGS}. The quadrupolar polar-type TLNs for more generic thin-shell gravastar models are presented in Ref.~\cite{Uchikata:2016qku}.

\subsubsection{On the universal BH limit}
It is remarkable that the models described above display a very similar behavior in the BH limit, when the radius $r_0\to 2M$, cf. Table~\ref{tab:summary}. Indeed, although all TLNs vanish in this limit, they have a mild logarithmic dependence.
On the light of our results, it is natural to conjecture that this logarithmic dependence is a generic feature of ultracompact exotic objects, and will hold true for any ECO whose exterior spacetime is arbitrarily close to that of a BH in the $r_0\to 2M$ limit.

Due to this mild dependence, the TLNs are not extremely small, as one would have naively expected if the scaling with $\xi$ were polynomial. Indeed, in the Planckian case ($r_0-2M\approx \ell_P$) the order of magnitude of the TLNs is the same for all models and it is given by Eq.~\eqref{estimates}. In particular, the TLNs of Planckian ECOs are only five orders of magnitude smaller than those a typical NS.
The detectability of these deviations from the ``zero-Love'' rule of BHs in GR is discussed in Sec.~\ref{sec:detectability}.

\section{Tidal perturbations of BHs beyond vacuum GR} \label{sec:BH}
In this section, we discuss the TLNs of BHs in other theories of gravity. Technical details are given in Appendix~\ref{app:comp}.

\subsection{Scalar-tensor theories}\label{sec:ST}
We start with scalar-tensor theories, which generically give rise to stationary BH solutions which
are identical to those of GR~\cite{hawking1972,Sotiriou:2011dz,Berti:2015itd}. Therefore, the background solution which we deal with is still
described by the Schwarzschild geometry. 
In the Jordan frame, neglecting the matter Lagrangian, the simplest example of scalar-tensor theory is described by the Brans-Dicke action (cf., e.g., Ref.~\cite{Berti:2015itd})
\begin{equation}
\label{bransdickeaction}
S=\frac{1}{16\pi}\int d^4 x\,\sqrt{-g} \left(\Phi R-\frac{\omega_\text{BD}}{\Phi}\partial_\mu\Phi\partial^\mu\Phi\right)\,,
\end{equation}
where $\omega_\text{BD}$ is a dimensionless coupling constant and $\Phi$ is a scalar field characteristic of the theory. Action~\eqref{bransdickeaction} yields the equations of motion,
\beq
&&G_{\mu\nu}=\frac{\omega_\text{BD}}{\Phi^2}\left(\partial_\mu\Phi\partial_\nu\Phi-\frac12 g_{\mu\nu}\partial_{\lambda}\Phi\partial^\lambda\Phi\right)+\frac{1}{\Phi}\nabla_\mu\nabla_{\nu}\Phi\,,\nn\\\label{BDeqmov1}\\
&&\square\Phi=0\,. \label{BDeqmov2}
\eeq
As mentioned above, the background solution is Schwarzschild with a vanishing scalar field.

Following the procedure described in Sec.~\ref{setup}, we consider
metric perturbations given by Eqs.~\eqref{heven} and~\eqref{hodd} for the polar and axial sector, respectively, and a scalar field perturbation given by Eqs.~\eqref{scalarfield} and~\eqref{eq:scalardecomp}. Since the scalar perturbations are even-parity, axial gravitational perturbations do not couple to them, implying that this sector is governed by equations identical to those of vacuum GR. Therefore, all axial-type TLNs of a nonrotating BH in Brans-Dicke gravity are zero, $k^B_{l}=0$.

On the other hand, in the polar sector, scalar perturbations can be obtained from Eq.~\eqref{BDeqmov2} by using the decomposition in Eqs.~\eqref{scalarfield} and~\eqref{eq:scalardecomp},
\begin{equation}
\delta\phi''+\frac{2 (r-M) \delta\phi'-l (l+1) \delta\phi}{r (r-2 M)}=0\,,
\end{equation}
The solution which is regular at the horizon is
\begin{equation}
\label{BDscalarfield}
\delta\phi= C_l P_l\left(\frac{r}{M}-1\right)\,,
\end{equation}
where $P_l$ is a Legendre polynomial, $C_l$ is an integration constant, and we have expanded the scalar field as in Eq.~\eqref{scalarfield} with $\Phi^{^{(0)}}=0$. By comparing the above expression with the scalar-field expansion in Eq.~\eqref{eq:scalarexpansion}, we conclude that $C_l\propto \mathcal{E}_l^{\rm S}$ and that the induced scalar multipoles $\Phi_l$ are zero. Therefore, Eq.~\eqref{BDscalarfield} represents an external scalar tidal field and the ``scalar TLN'' are identically zero.

Although we wish to focus on gravitational tidal fields, it is instructive to investigate the role of a scalar tide in scalar-tensor theory. By substituting Eq.~\eqref{BDscalarfield} in Eq.~\eqref{BDeqmov1} we obtain an inhomogeneous differential equation for $H_0$, which one of the polar perturbations of the metric [cf.\ Eq.~\eqref{heven}]. For $l=2$, we can identify $C_2\equiv -\frac{2}{3} M^2 \mathcal{E}_2^{\rm S}$ and we get
%
\begin{eqnarray}
H_0''&+&\frac{2 (r-M)}{r (r-2 M)}H_0'-\frac{2 \left(2 M^2-6 M r+3 r^2\right)}{r^2 (r-2 M)^2}H_0\nn\\
&=&\frac{4M^2 \left(2 M^2-6 M r+3 r^2\right)}{3r^2 (r-2 M)^2} \mathcal{E}_2^{\rm S}\,.\label{difeqH0BD}
\end{eqnarray}
%
The above equation can be solved analytically. The solution which is regular at the horizon reads
\begin{equation}
\label{solHBD}
H_0=-r^2 \mathcal{E}_2+2 M r \mathcal{E}_2-\frac{2}{3} M^2 \mathcal{E}^{\rm S}_2\,.
\end{equation}
The induced quadrupolar moment is zero, and therefore $k^E_{2}=0$, just as in the GR case.
It is straightforward to show that this result generalizes to higher multipoles, $k^E_l=0$.
In conclusion, although in Brans-Dicke theory the BH metric perturbations depend on scalar tides, all TLNs of a static BH vanish, as in the case of GR.

\subsection{Einstein-Maxwell} \label{sec:RNBH}
We consider Reissner-Nordstr\"om BHs, which are the unique static solution to Einstein-Maxwell theory, although our results are valid for any $U{(1)}$ field minimally coupled to gravity, as in the case of dark photons or the hidden $U{(1)}$ dark-matter sector~\cite{Cardoso:2016olt}. 
The Einstein-Maxwell field equations read
\begin{align}
\label{einstein}
&G_{\mu\nu}=8\pi \,T_{\mu\nu},\\
\label{maxwell}
&\nabla_\mu F^{\mu\nu}=0,
\end{align} 
where $F_{\mu\nu}=A_{\nu,\mu}-A_{\mu,\nu}$ is the Maxwell tensor and 
\begin{equation}
T_{\alpha\beta}=\frac{1}{4\pi}\left(g^{\mu\gamma}F_{\alpha\mu}F_{\beta\gamma}-\frac14 F_{\mu\nu}F^{\mu\nu}g_{\alpha\beta}\right)\,,
\end{equation}
is the stress-energy tensor of the EM field.
The background spacetime is the well-known Reissner-Nordstr\"om metric, whose line element reads as in Eq.~\eqref{staticmetric} with
\begin{equation}
\label{metricRN}
e^{\Gamma}=e^{-\Lambda_g}=1-\frac{2M}{r}+\frac{Q^2}{r^2}\equiv f(r),
\end{equation}
where $M$ and $Q$ denote the mass and the charge of the BH, respectively. The background Maxwell \mbox{$4$-potential} reads
\begin{equation}
A^{^{(0)}}_{\mu}=\left(-Q/r,0,0,0\right).
\end{equation}
Because the background is electrically charged, gravitational and EM perturbations are coupled to each other. To compute the tidal deformations, we expand the metric as in Eqs.~\eqref{heven} and \eqref{hodd} and the Maxwell field as in Eqs.~\eqref{4pot} and~\eqref{eq:EMpot}. As before, we consider the polar and the axial sectors separately.
\subsubsection{Polar TLNs}
The polar functions of the metric are coupled to the EM function $u_1$ through the field equations.
In the Lorenz gauge, we find the following coupled equations,
\begin{align}
\label{rn:difeq1}
&\mathcal{D}^{^{(2)}}_1 H_0 + \frac{4 Q}{r^3-2 M r^2+Q^2 r}\mathcal{D}^{^{(1)}}_1 u_1=0\,,\\
\label{rn:difeq2}
&\mathcal{D}^{^{(2)}}_2 u_1 + \frac{Q}{r}\mathcal{D}^{^{(1)}}_2 H_0=0\,,
\end{align}
where we defined the operators,
\beq
\mathcal{D}^{^{(2)}}_1&=&\frac{d^2}{dr^2} -\frac{2 (M-r)}{r^2 f} \frac{d}{dr}+\frac{1}{r^6 f^2}\left[Q^2 r (4 M-({\eta} -2) r)\right.\nn\\
&&\left.-r^2 \left(4 M^2-2 {\eta}  M r+{\eta}  r^2\right)-2 Q^4 \right]\,,\nonumber\\
\mathcal{D}^{^{(1)}}_1&=&\frac{d}{dr} + \frac{\left(Q^2-r^2\right)}{r \left(r (r-2 M)+Q^2\right)}\,,\nonumber\\
\mathcal{D}^{^{(2)}}_2&=&\frac{d^2}{dr^2} + \frac{4 Q^2-{\eta} r^2}{r^4 f}\frac{d}{dr} \,,\nonumber\\
\mathcal{D}^{^{(1)}}_2&=&\frac{d}{dr}+\frac{2 \left(M  r-Q^2\right)}{r^3f} \,,\nonumber
\eeq
with ${\eta}:=l(l+1)$. This system allows for a closed-form solution. For simplicity, we impose the absence of electric tidal fields, which requires that the function $u_1$ does not contain $r^3$-terms at large distance [cf.\ Eq.~\eqref{eq:Atexpansion}]. In this case, the regular solution at the horizon for $l=2$ reads
\beq
H_0^{l=2}&=&-\mathcal{E}_2 r^2f\,,\\
u_1^{l=2}&=&-\frac{\mathcal{E}_2 r^2Qf}{2}\,.
\eeq
Due to the gravito-EM coupling, an external tidal field induces a Maxwell perturbation which is proportional to the BH charge $Q$. A simple comparison between the above results and the expansions in Eqs.~\eqref{eq:gttexpansion} and~\eqref{eq:Atexpansion} shows that the multipole moments are all vanishing. 

Although the full solutions for $l>2$ are cumbersome, it can be shown that for any $l>2$ the large-distance expansion of the solutions for $H_0$ and $u_1$ which are regular at the horizon is truncated at the $1/r$ term for any $l$, and it is an exact solution of the coupled system. Therefore, the above result directly extends to any $l$, and we obtain that the TLNs of a charged BH are zero in the polar sector, $k^E_l=0$, like in the Schwarzschild case.
\subsubsection{Axial TLNs}
The calculations for gravitational axial TLNs and magnetic TLNs
are simpler. In this case we consider the axial sector of the  metric perturbations [cf.\ decomposition in Eq.~\eqref{hodd}] and of the Maxwell field [cf.\ Eq.~\eqref{eq:EMpot}].
The $r\varphi$-component of Einstein's
equations leads to $h_1=0$ which automatically satisfies the $\theta\varphi$-component.
The final axial system reads
\begin{eqnarray}
&&r^2f \left(r^2h_0''-\frac{4Qu_4'}{{\eta}}\right)-\left[2Q^2+r\left(r{\eta} -4M\right)\right]h_0=0 \,,\nn\\ \\
&&r^2fu_4''-{\eta} Qh_0'+2Mu_4'-{\eta} u_4-\frac{2Q}{r}(Qu_4'-{\eta} h_0)=0 \,. \nn\\
\end{eqnarray}
Also in the axial sector, the coupled system admits an analytic, closed-form solution. In the absence of EM tidal fields, the solutions which are regular at the horizon read
\beq
h_0^{l=2}&=&\frac{r^3}{3} f \mathcal{B}_2\,,\\
u_4^{l=2}&=&\frac{r^2}{2}Q \mathcal{B}_2(1-Q^2/r^2)\,.
\eeq
which proves that also the axial TLNs of a charged BH are zero. It is straightforward to extend this result to higher multipoles, finding
$k_l^{B}=0$.

To conclude, we obtain the interesting result that all TLNs or a static BH in Einstein-Maxwell theory are identically zero, as in the uncharged case.

\subsection{Chern-Simons gravity} \label{sec:Chern}

In this section we compute the TLNs of a nonrotating BH in Chern-Simons theory~\cite{Alexander:2009tp},
\begin{equation}\label{eq:actiondCS}
S_{\text{CS}}=\int d^4x\,\sqrt{-g}\left[R -\frac{1}{2}g^{ab}\partial_a\Phi\partial_b\Phi+\frac{\alpha_{\rm CS}}{4} \Phi \,{}^*\!RR \right],
\end{equation}
where $\alpha_{\rm CS}$ is the coupling constant of the theory and ${}^*RR$ is the Pontryagin scalar,
\begin{equation}
{}^*\!RR=\frac{1}{2}R_{abcd}\epsilon^{baef}{R^{cd}}_{ef}\,.
\end{equation}
The field equations arising from action~\eqref{eq:actiondCS} are
\begin{align}
\label{eq:dCSfieldeq1}
&R_{ab}= \frac{1}{2}\partial_a\Phi\partial_b\Phi-\alpha_{\rm CS} C_{ab} \,, \\
\label{eq:dCSfieldeq2}
&\square\Phi=-\frac{\alpha_{\rm CS}}{4}{}^*\!RR,
\end{align}
where $C_{ab}=\nabla_c\Phi\epsilon^{cde(a}\nabla_eR^{b)}_d+\nabla_{d}\nabla_{c}\Phi{}^*\!R^{dabc}$.
We will focus on spherically symmetric background solutions to Eqs.~\eqref{eq:dCSfieldeq1} and \eqref{eq:dCSfieldeq2}. In these conditions, the Pontryagin scalar vanishes and the background is described by the Schwarzschild metric with a vanishing scalar field~\cite{Jackiw:2003pm,Yunes:2007ss,Molina:2010fb}.

\subsubsection{Polar TLNs}

In Chern-Simons gravity, the polar perturbations of a Schwarzschild BH are equivalent to GR~\cite{Cardoso:2009pk,Molina:2010fb}. Therefore, the analysis for the polar TLNs is identical to that discussed in Ref.~\cite{Binnington:2009bb}, and one can conclude that all polar TLNs of a nonrotating BH in Chern-Simons gravity are zero, $k^E_{l}=0$.
As discussed in Sec.~\ref{sec:detectability}, the polar TLNs are the dominant correction to the inspiral waveform~\cite{Vines:2011ud}. Thus, the simple fact that in Chern-Simons gravity these TLNs are vanishing already suggests that it would be very difficult to constrain this theory with GW measurements of the BH tidal deformability.

\subsubsection{Axial TLNs}

On the other hand, the field $\Phi$ transforms as a pseudoscalar and is therefore part of the axial sector. We can thus expect that nontrivial axial-type TLNs in Chern-Simons gravity may exist.  
In the stationary limit, we find $h_1=0$, and the field equations for the axial sector reduce to a system of two coupled second-order differential equations for $h_0$ and $\delta\phi$. 
This system can be solved numerically for a generic coupling $\alpha_{\rm CS}$ or perturbatively when $\zeta_{\rm CS}:=\alpha_{\rm CS}/M^2\ll1$. The latter case is consistent with the action~\eqref{eq:actiondCS} being an effective field theory~\cite{Alexander:2009tp}. We have adopted both procedures, described below.

In the perturbative limit, we expand the metric and scalar perturbations in powers of the coupling $\zeta_{\rm CS}\ll 1$,
\beq
h_{\mu\nu}&=&h_{\mu\nu}^{(0)}+\zeta_{\rm CS}^2h_{\mu\nu}^{(2)}+...\\
\delta\phi&=&\zeta_{\rm CS}\delta\phi^{(1)}+...
\eeq
and solve the perturbation equations order by order in $\zeta_{\rm CS}$.

\paragraph{Quadrupolar TLNs.}

For $l=2$ and to ${\cal O}(\zeta_{\rm CS}^0)$, the metric perturbation which is regular at the horizon reads
\be
h_0^{(0)}=\frac{\mathcal{B}_2}{3}r^3\left(1-\frac{2M}{r}\right)\,,
\ee
as in the GR case.
The advantage of the perturbative approach is that the equations decouple from each other. 
To ${\cal O}(\zeta_{\rm CS}^1)$, the only correction is in the scalar-field equation, which reads
\begin{equation}
 {\cal D}_S^{(2)}\delta\phi^{(1)} = \frac{12\mathcal{B}_2 M}{r^2(r-2M)}\,,
\end{equation}
where
\begin{equation}
{\cal D}_S^{(l)}:= \frac{d^2}{dr^2}-\frac{2(M-r)}{r^2-2Mr}\,\frac{d}{dr}-\frac{l(l+1)}{r^2-2Mr}\,.
\end{equation}
Again for simplicity, we impose the absence of scalar tidal fields, i.e. we require that $\delta\phi$ does not contain a divergent $r^l$ term at large distance [cf. Eq.~\eqref{eq:scalarexpansion}]. In this case, the solution which is regular at $r=2M$ reads
%
\begin{eqnarray}
\delta\phi^{(1)} &=& -\frac{\mathcal{B}_2 M^2}{2} \biggl(
54-36y+\pi ^2 \left(2+3 (y-2) y\right) \nn\\
&+&3\log\left[\frac{y}{2}\right] \left[12 (1-y)+(2+3 (y-2) y)\log\left[\frac{y}{2}\right]\right]\nn\\
&+&6(2+3 (y-2)y) \polylog\left[2,1-\frac{y}{2}\right] \biggr)\ ,
\end{eqnarray}
%
where we defined $y=r/M$.
With the above solution at hand, the ${\cal O}(\zeta_{\rm CS}^2)$ equation for the axial perturbation reads ${\cal D}_{A}^{(2)}{h_0^{(2)}} = \mathcal{S}_A^{(2)}$ with
%
\begin{eqnarray}
\mathcal{S}_A^{(2)}(r)&:=& \frac{3\mathcal{B}_2 M}{(y-2) y^4} \biggl((2-y) 
\left[\pi ^2 \left(3 y^2-2\right)-18 (1+2 y)\right]\nn\\
&-&6 (y-2) \left(3 y^2-2\right)  \polylog\left[2,1-\frac{y}{2}\right]+3 \log\left[\frac{y}{2}\right] \nn\\
&\times&\left[4 (3 (y-1) y-4)-(y-2) \left(3 y^2-2\right) \log\left[\frac{y}{2}\right]\right]\biggr) \,. \nn\\\label{DCSscalar}
\end{eqnarray}
%

\begin{figure}[th]
\includegraphics[width=0.45\textwidth]{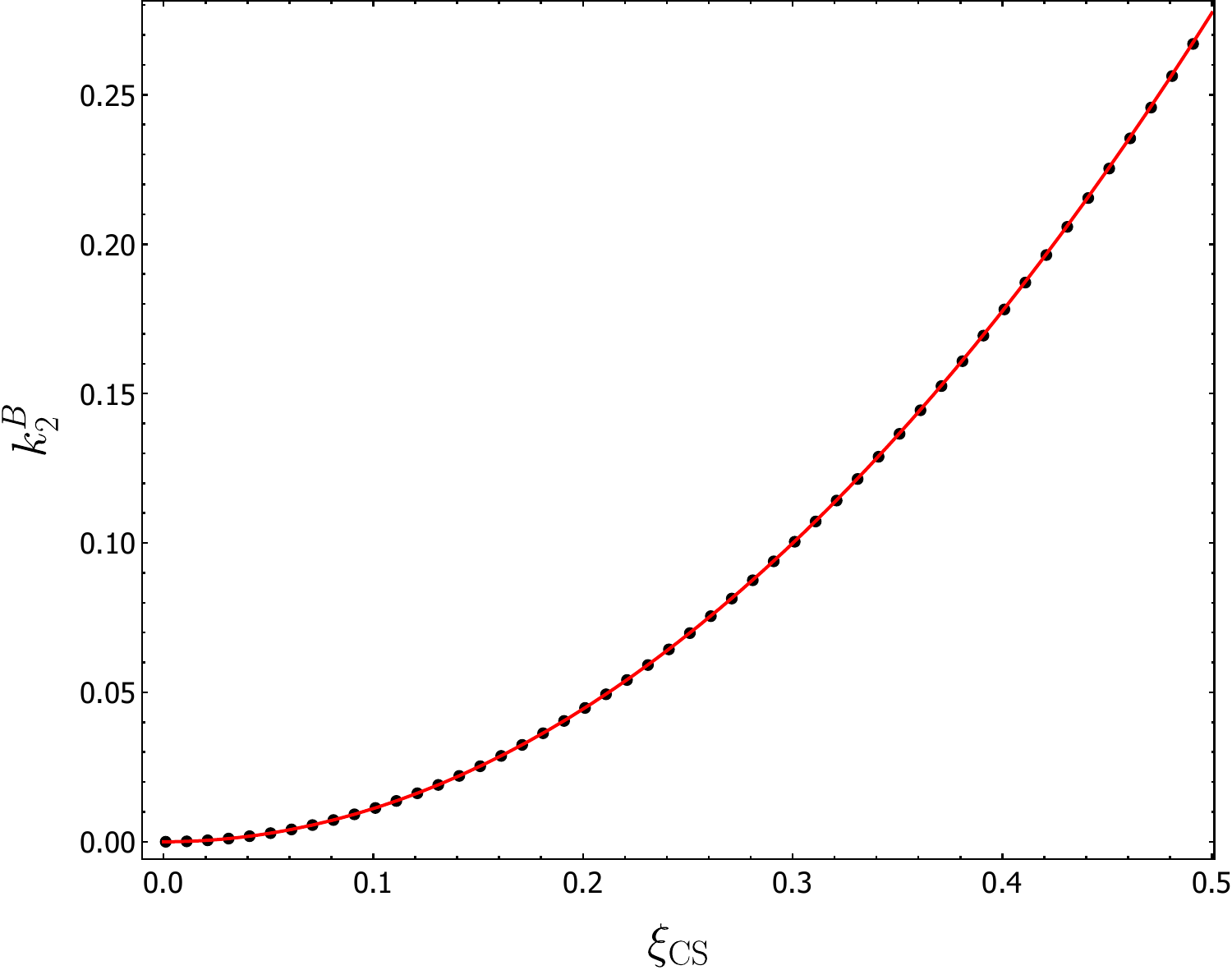}
\caption{Axial TLNs of a BH in Chern-Simons gravity for $l=2$ perturbations calculated for different values of $\alpha_{\rm CS}$. The dots correspond to the values obtained directly from a numerical integration, whereas the line is the analytical result shown in Eq.~\eqref{DCSl2}.}\label{fig:k2dCSalpha}
\end{figure}

It is convenient to solve the above inhomogeneous equation through the Green's function. The solution with the correct boundary conditions is
\begin{eqnarray}
 {h_0^{(2)}}(r) &=& \frac{\Psi_+(r)}{W}\int_{2M}^r dr' \mathcal{S}_A^{(2)}(r') \Psi_-(r')\nn\\
 &+&\frac{\Psi_-(r)}{W}\int_{r}^{\infty} dr' \mathcal{S}_A^{(2)}(r') \Psi_+(r') \,, \label{Green_CS}
\end{eqnarray}
where the two linearly independent solutions of the homogeneous problem read
\begin{eqnarray}
\Psi_-(r) &=& \frac{A_1 r^2 (r-2 M)}{4 M^3}\,,\\
\Psi_+(r) &=& \frac{A_2}{6 M^3 r} \biggl(2 M \left(2 M^3+2 M^2 r+3 M r^2-3 r^3\right) \nn\\
&+&3 (2 M-r) r^3 \log(1-2M/r)\biggr)\,,
\end{eqnarray}
and $W= A_1 A_2/M$ is the Wronskian, which is constant by virtue of the field equations. The above solutions are regular at the horizon and at infinity, respectively.
At large distances, $\mathcal{S}_A^{(2)}\sim \mathcal{B}_2/r^5$, and the first integral in Eq.~\eqref{Green_CS} is convergent, whereas the second integral does not contribute to the current quadrupole moment $S_2$. Interestingly, these integrals can be computed in closed form although their final expression is cumbersome. We report here only the large-distance behavior of ${h_0^{(2)}}(r)$, namely
\begin{equation}
 {h_0^{(2)}}(r) \to \frac{9}{5} [9-8 \zeta(3)]\frac{\mathcal{B}_2 M^5 }{r^2}+{\cal O}\left(\frac{M^3}{r^3}\right)\,,
\end{equation}
where $\zeta(n)$ is the Riemann Zeta function. By comparing the above result with Eq.~\eqref{eq:gtphiexpansion} and using Eq.~\eqref{Lovenumbersdef1}, it is straightforward to obtain
\begin{equation}
 k_2^{\rm B} = \frac{9}{5} [8 \zeta (3)-9]\zeta_{\rm CS}^2\approx 1.10962\,\zeta_{\rm CS}^2 \,. \label{DCSl2}
\end{equation}
Interestingly, we find that the axial TLN is nonzero and proportional to $\zeta_{\rm CS}^2$, as expected. We have also confirmed this result by integrating numerically the field equations for arbitrary values of $\alpha_{\rm CS}$ and by extracting the quadratic correction in the small-coupling limit.
Figure~\ref{fig:k2dCSalpha} shows a comparison between the analytical result~\eqref{DCSl2} and the numerical one.

\paragraph{Octupolar TLNs.}

Although the computation for the $l=3$ case proceeds as presented above for $l=2$, there are some notable differences. In this case, the ${\cal O}(\zeta_{\rm CS}^0)$ perturbation reads
%
\be
h_0^{(0)} =\frac{\mathcal{B}_3 M}{27} r^2(r-2M)  (3 r-4M)\,,
\ee
whereas the scalar-field equation to ${\cal O}(\zeta_{\rm CS})$ is
\begin{equation}
 {\cal D}_S^{(3)}\delta\phi^{(1)} =  \frac{16 \mathcal{B}_3 M^3(3r-5M)}{3 (r-2M) r^2}\,,
\end{equation}
and its solution regular at the horizon can be written in a form similar to Eq.~\eqref{DCSscalar}. However, even after imposing the absence of scalar tides, the large-distance behavior of the scalar field is $\delta\phi^{(1)}\sim 4 \mathcal{B}_3/3 +{\cal O}(M/r)$, i.e. it approaches a constant value. This has important consequences for the metric perturbations to ${\cal O}(\zeta_{\rm CS}^2)$, which are governed by the equation ${\cal D}_{A}^{(3)}{h_0^{(2)}} = \mathcal{S}_A^{(3)}$, where now the source term $\mathcal{S}_A^{(3)}\sim 8 \mathcal{B}_3/r^{-4}$ at large distances. Because the divergent solution of the homogeneous problem reads $\Psi_-\to r^4$ at large distances, the first integral in Eq.~\eqref{Green_CS} (whose schematic form is valid for any values of $l$) yields terms which grow linearly and logarithmically at large distances. By repeating the same procedure as before, we obtain the following large-distance behavior for $l=3$,
\begin{eqnarray}
h_0^{(2)} &\sim& -\frac{4 \mathcal{B}_3M^6}{3 r^2}-\frac{4\mathcal{B}_3M^7}{3969 r^3} \biggl(75600 \zeta (3)-84331 \nn\\
&-&7560  \log\left(\frac{r}{2M}\right)\biggr)+{\cal O}\left(\frac{M^4}{r^4}\right)\,.
\end{eqnarray}
We therefore obtain two terms (the $1/r^2$ and the $\log r/r^3$ term) which decay more slowly than the octupole term, $1/r^3$.
We believe that these terms arise as subleading corrections to the external tidal field, which is not captured by the asymptotic expansion in Eq.~\eqref{eq:gtphiexpansion}, because the latter does not include the effects of a (nonminimally coupled) scalar field at infinity.
In other words, the multipolar structure of a tidally deformed BH in Chern-Simons gravity is more involved than in GR.
We anticipate that this issue also appears in other modified theories and we postpone a more detailed analysis to the future~\cite{inprep}. In order to get an estimate for the TLN, we simply consider the ordinary octupolar correction. By proceeding in the usual way, we obtain
\begin{equation}
 k_3^{\rm B} = \frac{1}{588} \left(75600 \zeta (3)-84331\right)\zeta_{\rm CS}^2\approx 11.13\,\zeta_{\rm CS}^2 \,, \label{DCSl3}
\end{equation}
which is the value reported in Table~\ref{tab:summary} although, for the reasons mentioned above, should only be considered as an estimate.

\section{Detectability} \label{sec:detectability}
To estimate the detectability of the TLNs through GW observations, we use a Fisher matrix approach.  
In the following, we summarize the basic features of this formalism, referring the reader 
to Ref.~\cite{Vallisneri:2007ev} (and to the references therein) for a detailed discussion of the topic.

We consider the output of a generic interferometer,
\begin{equation}
s(t)=h(t,\vec{\xi})+n(t)\ ,
\end{equation}
where $n(t)$ is the detector noise (assumed stationary), and $h(t,\vec{\xi})$ is the GW signal. 
Our goal is to recover the physical parameters $\vec{\xi}=\{\xi_1,\ldots,\xi_n\}$, and their 
errors $\Delta\vec{\xi}=\vec{\xi} -\vec{\chi}$ with respect to the true values $\vec{\chi}$. 
We therefore need to compute the  
probability distribution $p(\vec{\xi}\vert s)$, which can be written as
\begin{equation}\label{prob}
p(\vec{\xi}\vert s)\propto p^{(0)}(\vec{\xi})e^{-\frac{1}{2}(h(\vec{\xi})
-s\vert h(\vec{\xi})-s)}\ ,
\end{equation}
with $p^{(0)}(\vec{\xi})$ prior on the parameters~\cite{PhysRevD.49.2658}.
The bracket $(\cdot \vert \cdot)$ represents the inner product
\begin{equation}\label{inner}
(g\vert h)=2\int_{-\infty}^{\infty}df\,\frac{\tilde h(f)\tilde g^{*}(f)+\tilde h^{*}(f)\tilde g(f)}{S_{h}(f)}\ ,
\end{equation}
where $S_h(f)$ is the detector's noise spectral density and $\tilde{h}(f)$ is the Fourier transform 
of the signal.
According to the principle of the maximum-likelihood estimator, the values of the
source parameters can be estimated as those which maximize Eq.~(\ref{prob}).  
In the limit of large signal-to-noise ratio, such that $p(\vec{\xi}\vert h)$ is tightly 
peaked around the true values of the source parameters, a Taylor expansion of 
$p(\vec{\xi}\vert s)$ around $\vec{\chi}$ leads to
\begin{equation}\label{prob1}
p( \vec{\xi}\vert s)\propto p^{(0)}(\vec{\xi})e^{-\frac{1}{2}\Gamma_{ab}\Delta \xi^{a}\Delta \xi^{b}}\ ,
\end{equation}
where 
\begin{equation}\label{FisherM}
\Gamma_{ab}=\left(\frac{\partial h}{\partial \xi^a}\bigg\vert \frac{\partial h}{\partial \xi^b}\right)_{\vec{\xi}=\vec{\chi}}
\end{equation}
is the Fisher information matrix. Inversion of the latter yields the 
covariance matrix,
\begin{equation}\label{covariance}
\Sigma^{ab}=\left(\Gamma^{-1}\right)^{ab}\ .
\end{equation}
The error on the source parameters $\xi^{a}$ are then given by
\begin{equation}\label{error}
\sigma_{a}=\sqrt{\Sigma^{aa}}\ ,
\end{equation}
and the correlation coefficients between $\xi^{a}$ and $\xi^{b}$ are given by
\begin{equation}\label{corr}
c_{ab}=\frac{\langle \Delta \xi^{a}\Delta \xi^{b}\rangle}{\Sigma^{aa}\Sigma^{bb}}=
\frac{\Sigma^{ab}}{\sqrt{\Sigma^{aa}\Sigma^{bb}}}\ .
\end{equation}

The TLNs enter the GW signal as a fifth-order post-Newtonian (PN) correction 
which adds linearly to the phase of the waveform,
\begin{equation}
\tilde{h}(f)={\cal A}(f)e^{i(\psi_\tn{PP}+\psi_\tn{T})}\ ,
\end{equation}
where $\psi_\tn{PP}(f)$ is the point-particle contribution, while $\psi_\tn{T}(f)$ describes the tidal 
effects and --~to the leading order~-- depends on the $l=2$ polar TLNs through the constant~\cite{Flanagan:2007ix,Hinderer:2007mb}
\begin{equation}
 \lambda:=\frac{2}{3}M^5 k_2^E\,. \label{deflambda}
\end{equation}
The contribution of higher multipoles and of the axial-type TLNs is subleading and will be neglected in the following.

In our analysis, we use the so-called TaylorF2 
approximant of the GW template in the frequency domain~\cite{PhysRevD.62.084036}, which is 3.5PN accurate in 
the point-particle phase and 2PN accurate in the tidal term~\cite{PhysRevD.83.084051,PhysRevD.85.124034}.\footnote{For the purpose of this paper, we consider the amplitude at the leading order.}
For binary systems for which $\lambda_{i=1,2}\neq 0$, finite-size effects are described in terms of the average deformability,
\begin{equation}
\Lambda=\frac{1}{26}\left[\left(1+\frac{12}{q}\right)\lambda_1+(1+12 q)\lambda_{2}\right]\,, \label{defLambda}
\end{equation}
where $q:=m_1/m_2\geqslant 1$ is the mass ratio. For nonspinning objects, the sky-averaged waveform depends on 6 
parameters $\vec{\xi}=(\ln{\cal A},\phi_c,t_c,\ln {\cal M},\ln\nu,\Lambda)$, i.e.  
the amplitude, the phase and time at the coalescence, the chirp mass ${\cal M}=\nu^{3/5}(m_1+m_2)$, 
the symmetric mass ratio $\nu={m_1m_2}/{(m_1+m_2)^2}$ and the average tidal deformability defined in Eq.~\eqref{defLambda}. We also assume that the binary system follows circular orbits.
Nonetheless, $\ln {\cal A}$ is completely uncorrelated with the other variables, and therefore we will 
restrict our analysis by performing derivatives only with respect to the remaining parameters, leading 
to a $5\times 5$ Fisher matrix. 

The detector properties are encoded in the noise spectral density $S_h(f)$. We perform the analysis both for terrestrial and space interferometers.\footnote{In our numerical codes, we have taken into account the angle average of the pattern functions that depend on the orientation of the detector and the direction of the source. This introduces a corrective factor for the GW amplitude. Furthermore, in the case of LISA we have also included a geometrical correction factor $\sqrt{3}/2$ in the amplitude, to account for the fact that the LISA arms form a 60-degree angle~\cite{Cutler:1997ta,Berti:2005ys}. Finally, since we sky-average the signal, we will use an effective nonsky-averaged noise power spectral density, obtained by multiplying
$S_h$ by $3/20$ (see Ref~\cite{Berti:2004bd} for details).} 

For the Earth-based detectors, we consider (i)~AdLIGO with its anticipated design sensitivity curve 
\texttt{ZERO\_DET\_high\_P}~\cite{zerodet} and (ii)~the ET design configuration, with noise described by the analytic fit 
provided in Ref.~\cite{Sathyaprakash:2009xs}. As a space-based detector, we consider the most optimistic LISA configuration, namely the \texttt{N2A5} model defined in Ref.~\cite{PhysRevD.93.024003}, with a $5\times 10^6\,{\rm km}$ arm-length and an observing time of $T_\tn{obs}=5\,{\rm yr}$.

To compute the errors on the tidal deformability, we numerically integrate Eq.~\eqref{FisherM} within 
the frequency range $[f_\tn{min},f_\tn{max}]$, where $f^\tn{AdLIGO}_\tn{min}=20\,{\rm Hz}$, 
$f^\tn{ET}_\tn{min}=1\,{\rm Hz}$, and $f^\tn{LISA}_\tn{min}=\tn{Max}[10^{-5}\,{\rm Hz},4.149\times 10^{-5}
(10^{-6}{\cal M}/M_\odot)^{-5/8}(T_\tn{obs}/{\rm yr})^{-3/8}\,{\rm Hz}]$~\cite{PhysRevD.71.084025}. 
For the upper value we choose the frequency at innermost circular orbit for AdLIGO and ET, 
namely $f^\tn{AdLIGO}_{\tn{max}}=f^\tn{ET}_{\tn{max}}=(6^{3/2} m \pi)^{-1}$,
while for LISA $f^\tn{LISA}_\tn{max}=\tn{Min}[1\,{\rm Hz},(6^{3/2} m \pi)^{-1}]$, being $m=m_1+m_2$ the 
total mass of the system. For inspiral-merger-ringdown signals as those produced by binary coalescence, a sharp cutoff $f\leq f_\tn{max}$ which abruptly terminates the GW template does not alter the parameter covariance~\cite{Mandel:2014tca}.

\subsection{Model-independent tests with GWs}
%
\begin{figure*}[th]
\centering
\includegraphics[width=0.32\textwidth]{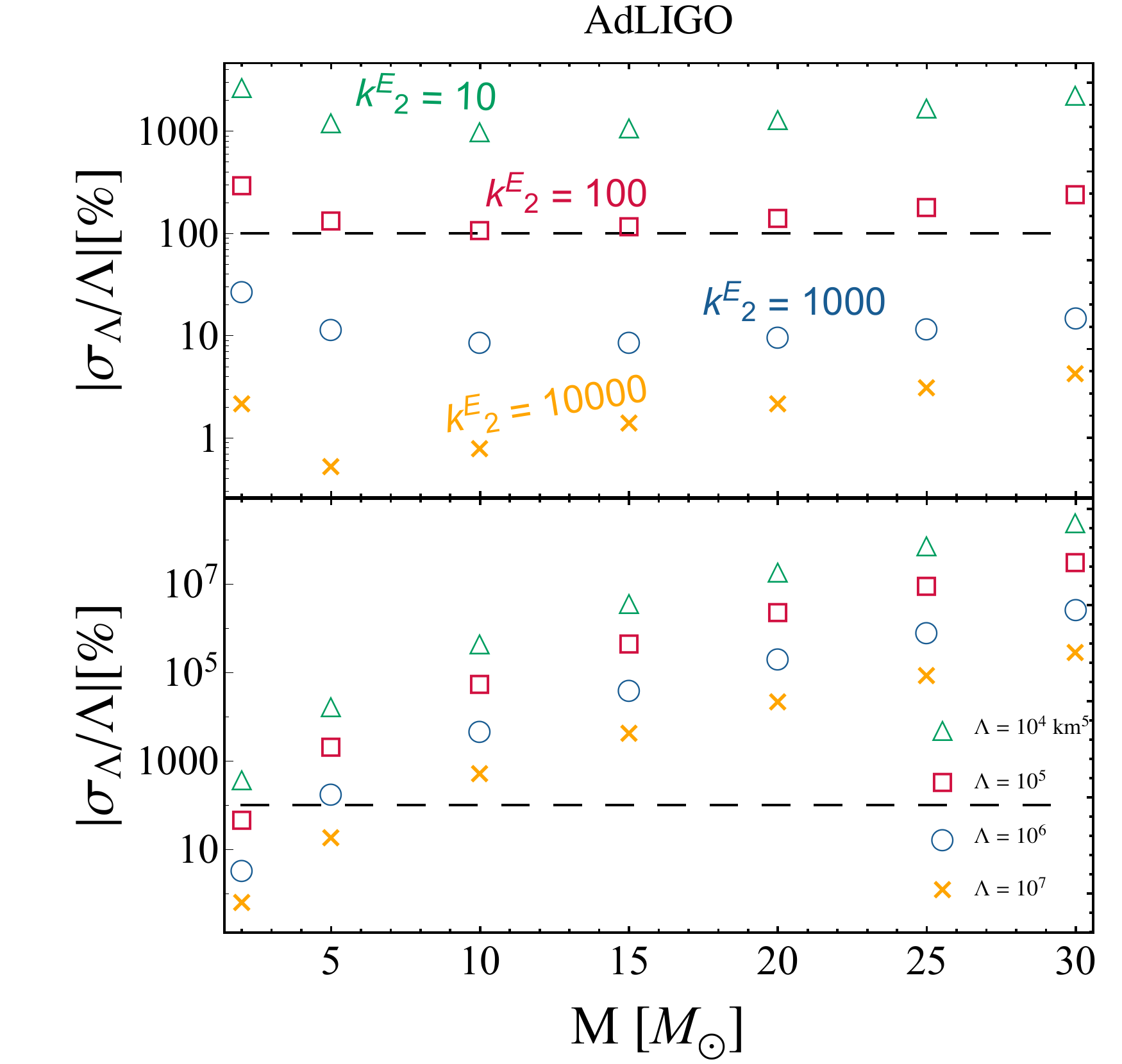}
\includegraphics[width=0.32\textwidth]{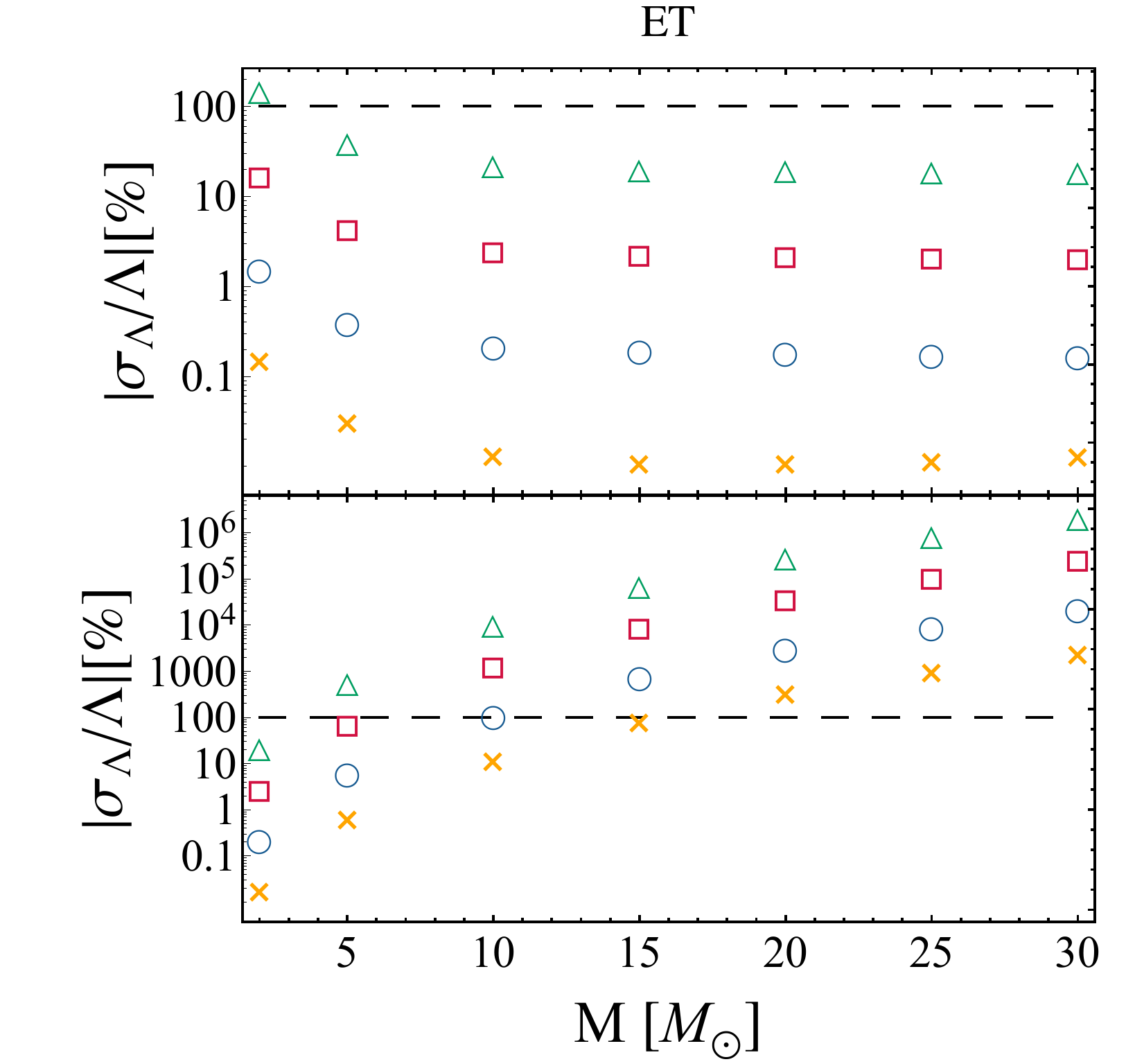}
\includegraphics[width=0.32\textwidth]{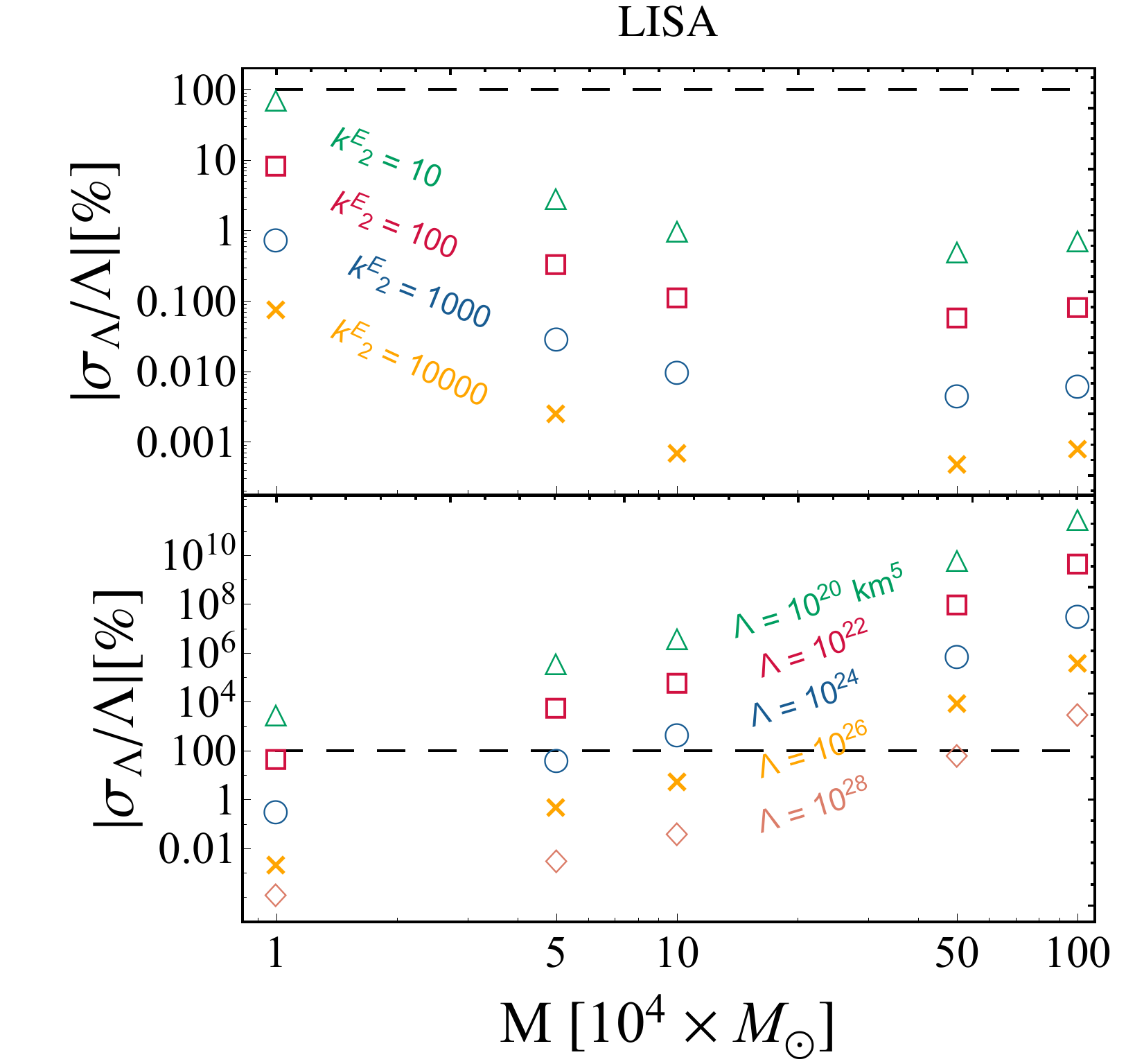}
\caption{Relative percentage errors on the average tidal deformability $\Lambda$ for equal-mass binaries at $100\,{\rm Mpc}$ (for AdLIGO and ET, left and middle panel, respectively) and at $500\,{\rm Mpc}$ (for LISA, right panel) as functions of the mass of the single object and for different values of the TLN $k_2^E$ (top panels) and of $\Lambda$ (bottom panels) of the two objects. The horizontal dashed line identifies the upper 
bound $\sigma_\Lambda/\Lambda=1$. The SNR corresponding to the considered configurations ranges between  $\sim[19\div 150]$ for AdLIGO, $\sim[265\div 1860]$ for ET, and $\sim[840\div 1.7\times 10^4]$ for LISA.}
\label{fig:detectabilitygeneric}
\end{figure*}

Before discussing the detectability for different families of ECOs, it is instructive 
to analyze the impact of the TLNs on the GW signal in a more general framework. 
Figure~\ref{fig:detectabilitygeneric} shows the relative uncertainty $\sigma_\Lambda/\Lambda$ 
for equal-mass binaries at $d=100\,{\rm Mpc}$ (for AdLIGO and ET) and at $d=500\,{\rm Mpc}$ (for LISA), as a function of the mass of the objects and for different values of the TLNs $k_2^E$ (top panels) and of the average tidal deformability $\Lambda$ (bottom panels).
In the panels of Fig.~\ref{fig:detectabilitygeneric}, the dashed horizontal line denotes the upper bound $\sigma_\Lambda=\Lambda$. Therefore, each point above that line is indistinguishable from a BH-BH binary in GR $(\Lambda=k_2^E=0$) within the errors, whereas a measurement of the TLNs for systems which lie below the threshold line would be incompatible to zero and, therefore, the ECOs can be distinguished from BHs in this case.

It is worth remarking that --~motivated by the prospect of measuring the TLNs of NSs through GW detections~-- several efforts have been devoted to investigate the detectability of $\Lambda$ for objects with $M\lesssim 2M_\odot$. The latter represents the mass range in which terrestrial interferometers will provide new information on matter at supranuclear densities from 
neutron-star binaries. On the other hand, our results shown in Fig.~\ref{fig:detectabilitygeneric} do not assume any specific model and extend the analysis of the detectability of the TLNs to a regime unexplored so far, where more massive ECOs can contribute to the GW signal through 
finite-size effects. Likewise, to the best of our knowledge, this work presents the first analysis on the detectability of tidal effects with LISA.

From the bottom panels of Fig.~\ref{fig:detectabilitygeneric} we note that, for a fixed $\Lambda$, the detectability is favored for low-mass systems, as the 
tidal phase scales with the inverse of the total mass $\psi_\tn{T}\propto \Lambda m^{-10/3}(1+q)^2/q$. 
Moreover, for $2M_\odot\lesssim M\lesssim 5M_\odot$, 
AdLIGO will constrain the TLNs for small compactness only (i.e., for large $\Lambda$). 
This picture improves for ET, which leads to an upper bound $\sigma_\Lambda/\Lambda=1$ 
up to $M\simeq 15 M_\odot$. Therefore, as far as terrestrial interferometers are considered, the high-compactness 
regime for ECOs seems to be available only for the third generation of detectors.
This result is also evident from the top panels of the left and middle plots in Fig.~\ref{fig:detectabilitygeneric}, which show that AdLIGO will not be able to set any significant constraint below $k_2^E\simeq 100$, regardless the ECO mass.

On the other hand, space interferometers open a completely new window onto finite-size effects. The top-right  panel of Fig.~\ref{fig:detectabilitygeneric} shows that  LISA is capable to 
bound the Love numbers with a relative accuracy $\sigma_\Lambda/\Lambda\lesssim 10\%$ in almost the entire mass range $M\in[10^4,10^6]\,M_\odot$. In other words, binary systems made of 
intermediate-mass compact objects will provide interesting constraints on the TLNs, with 
$k_2^E\simeq10$ and above, and therefore also on the nature of these objects. The exquisite precision 
of LISA can be traced back on the values of the SNR which characterize the massive binaries considered, 
which are up to 2 orders of magnitude larger than those obtained for terrestrial interferometers 
(see also caption of Fig.~\ref{fig:detectabilitygeneric}).


\subsection{Detectability of ECOs}

Let us now turn our attention to some specific models and, in particular, to the models of ECOs investigated in the previous sections. 
Based on the previous discussion, as a general setup we consider equal-mass binaries at 
distances $d=100\,{\rm Mpc}$ with $M\in[2,30]M_{\odot}$ for AdLIGO/ET, and at $d=500\,{\rm Mpc}$ with $M\in[10^4,10^6]M_{\odot}$ 
for LISA. We note that the GW signal is proportional to $1/d$, and therefore the covariance 
matrix~\eqref{covariance} (i.e., the error on $\Lambda$) scales linearly with the distance.

\begin{figure*}[th]
\centering
\includegraphics[width=0.32\textwidth]{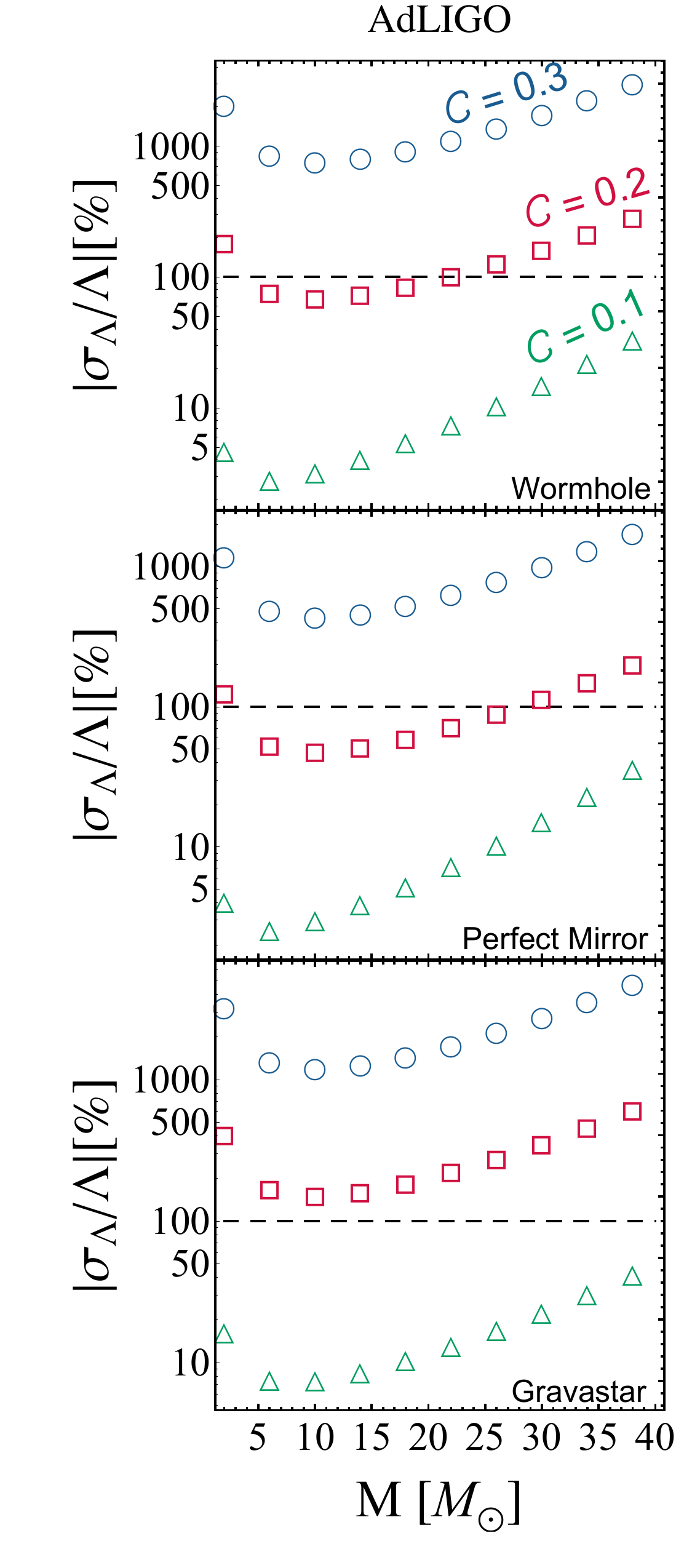}
\includegraphics[width=0.32\textwidth]{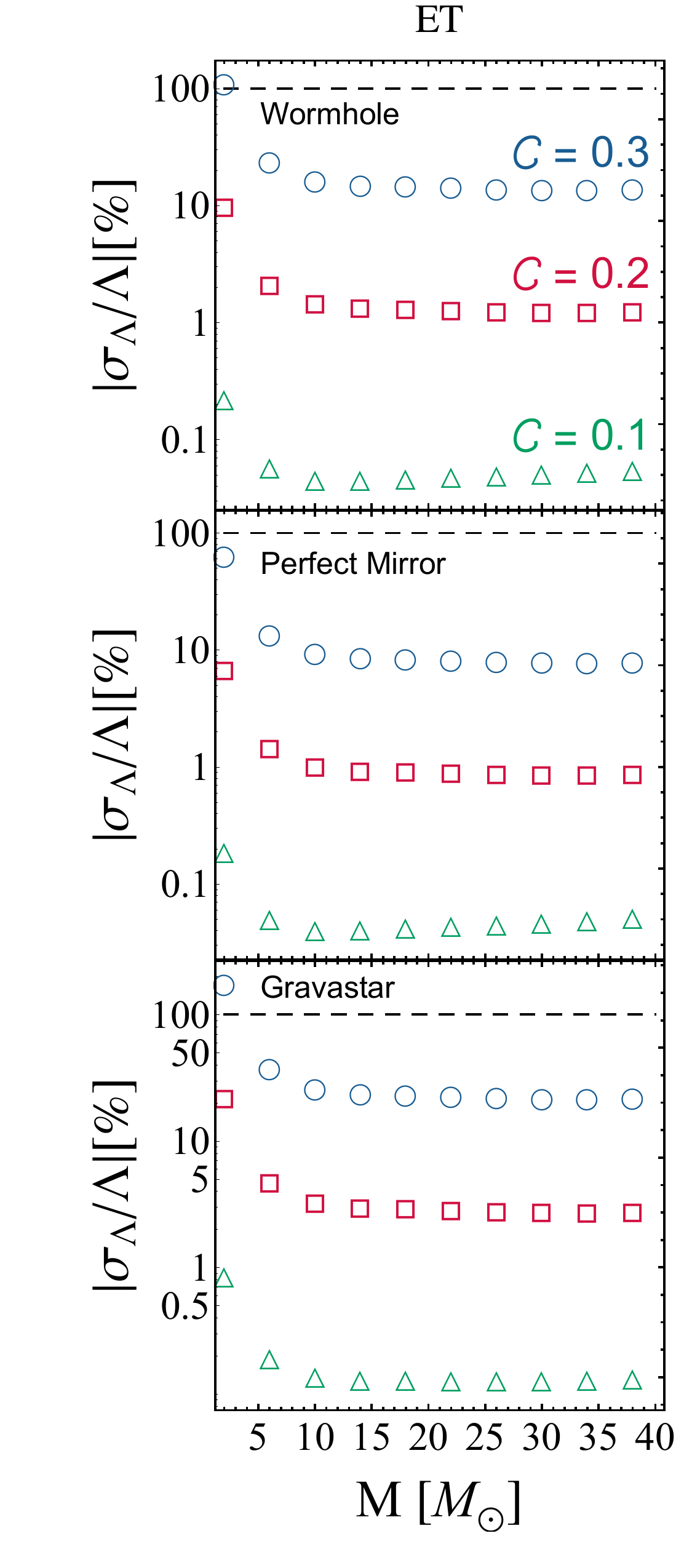}
\includegraphics[width=0.32\textwidth]{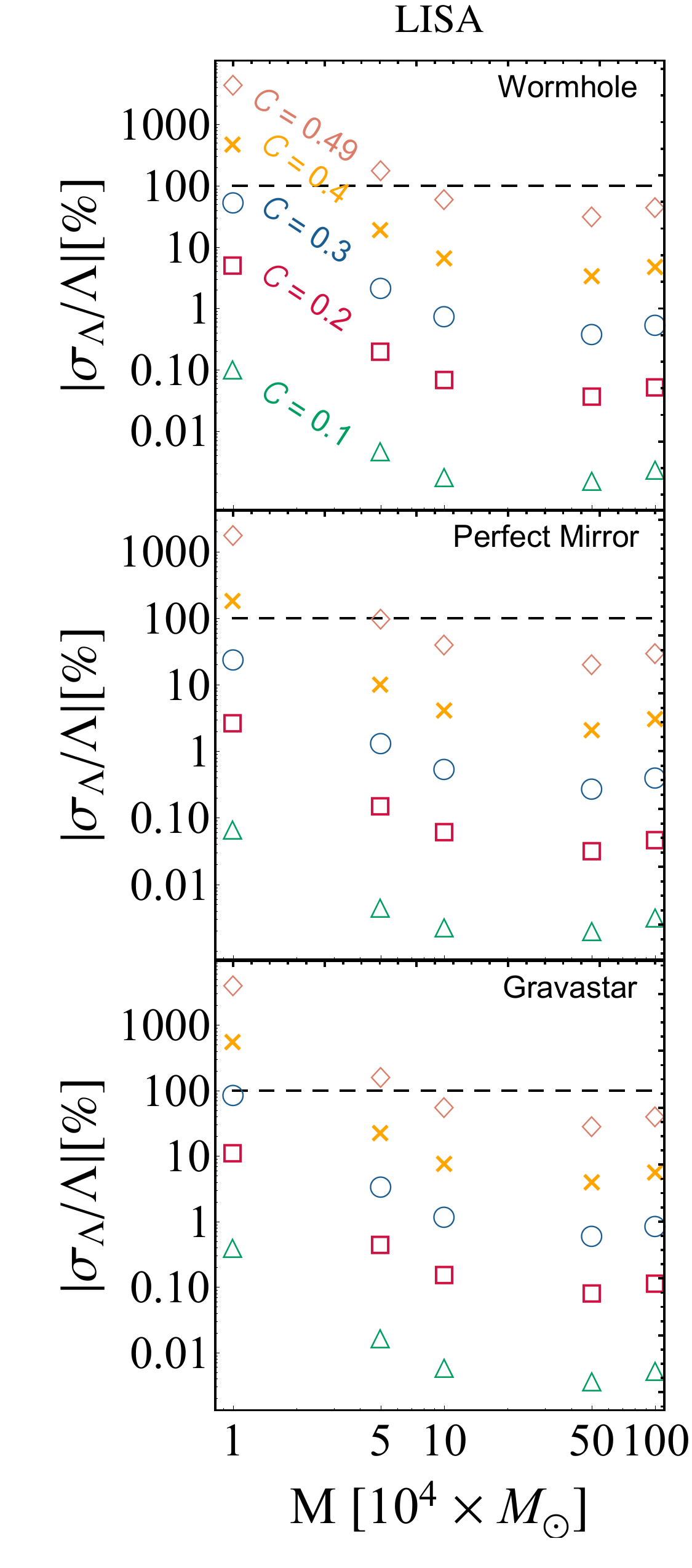}
\caption{Relative percentage errors on the tidal deformability for binaries observed by AdLIGO (left panels), 
ET (middle panels), and LISA (right panels), as functions of the ECO mass and for different values of the compactness. For terrestrial 
interferometers we consider prototype binaries at $d=100\,{\rm Mpc}$, while for LISA we set the source at $d=500\,{\rm Mpc}$.
Top, middle and bottom panels refer to wormholes, perfect-mirror models, and gravastars, respectively.
}
\label{fig:detectability}
\end{figure*}

In Fig.~\ref{fig:detectability} we show the percentage relative errors $\sigma_\Lambda/\Lambda$ for 
models of wormholes, perfect mirrors, and gravastars, as a function of the mass of the object and for different values of its 
compactness. 
Some qualitative results are independent of the nature of the ECO: the left panels confirm that AdLIGO would be able to constrain 
the tidal deformability only for small values of the compactness, namely ${C}\lesssim 0.2$. As the errors 
scale with the distance, an upper bound $\sigma_\Lambda/\Lambda\sim 1$ for ${C}=0.3$,
would require a source located at a distance $\sim 10\,{\rm Mpc}$. 

Furthermore, the relative errors decrease for 
larger masses, reach a minimum, and then increase again. This behavior can be explained by looking at the 
functional form of Eq.~\eqref{deflambda}. For a fixed compactness (i.e., for fixed $k_2^E$), the average tidal deformability grows  
with the ECO mass, thus making the tidal part of the gravitational waveform more easy to be detected. However, 
since the template is truncated at the last stable orbit, $f_\tn{max}\sim m^{-1}$, increasing the mass also reduces 
the number of effective cycles spent into the detector's bandwidth. This is particular penalizing for tidal effects, which 
enter the GW signal as high-PN/high-frequency corrections. 

It is worth noticing that a network of advanced interferometers would 
improve these results, even though it will not drastically change the upper bound on the compactness of these objects.
Indeed, if we consider that the experiments are all independent, the Fisher matrices computed for each detector 
simply sum up, and the overall error on $\Lambda$ is given by the inverse of the total $\Gamma_{ab}$.
Assuming five detectors with the same sensitivity of AdLIGO, the relative error $\sigma_\Lambda$ would decrease roughly by a factor $\sqrt{5}$ which, from our results in Fig.~\ref{fig:detectability}, is still not enough to constrain objects much more compact than ${C}\sim0.2$.

Third-generation ground-based detectors, like ET (middle panels of Fig.~\ref{fig:detectability}), hold more promising results. In this case, the relative errors $\sigma_\Lambda/\Lambda$ decrease almost by two orders of magnitude relative to AdLIGO. A GW detection of an ECO binary at $d=100\,{\rm Mpc}$ 
would allow to distinguish the system from a BH-BH binary (by the sole detection of the TLNs) approximately up to compactness ${C}\sim0.35$.
 
This scenario improves drastically for space-based detectors such as LISA (right panels of Fig.~\ref{fig:detectability}). Within the considered mass range, tidal effects may be measured for ECOs with ${C}\lesssim 0.3$ up to $1\%$ of accuracy. 
Moreover, LISA will be able to put strong constraints even for more compact objects: for $M\gtrsim 10^5M_\odot$ 
it would be possible to set an upper bound $\sigma_\Lambda/\Lambda=1$ in the entire parameter space. As discussed in the previous 
section, these results rely on the magnitude of the ECO's mass, which strengthens the effect of tidal interactions 
in the waveform. The right panels of Fig.~\ref{fig:detectability} show indeed that for all the considered ECO models, 
LISA leads the analysis to nearly explore the BH limit ${\cal C}\rightarrow 1/2$.

It is worth remarking that, as finite-size effects develop during the late inspiral, eventually leading to complex 
phenomena like the excitations of modes~\cite{Steinhoff:2016rfi}, a more accurate 
template which extends the frequency domain of the waveform up to the merger phase, 
would improve this analysis.

\subsection{Detectability of BSs}

For each model of BSs listed in Table~\ref{tab:BSs}, we focus on the most compact configuration in the stable branch. The mass of this configuration depends on the parameters of the potential. In Fig.~\ref{fig:detectabilityBS}, we show the results of the Fisher matrix analysis for an equal-mass BS-BS binary as a function of the BS mass, obtained by considering the most compact configuration and by varying the parameters of the potential. Since, for each model, we consider the maximum compactness allowed in the nonspinning case, our results can be seen as conservative, since less compact configurations are easier to discriminate.

The forecast for detecting BS-BS binaries by using GW tidal effects are more promising than for other ECOs, because the compactness of BSs is at most $C\lesssim 0.3$ and, in turn, their TLNs are larger. Advanced LIGO will be able to discriminate between minimal BSs and BHs in all the mass range. These errors 
worsen for massive objects. However, for all the models analyzed, second-generation detectors will set a strong 
upper bound on the tidal deformability, unless solitonic BS are considered. 
Results in Fig.~\ref{fig:detectabilityBS} suggest that models of minimal and massive BSs can be distinguished from BHs through AdLIGO detections, whereas discriminating between BHs and more compact BSs such as solitonic models will require future detectors like ET. 
Again, the uncertainties significantly decrease with LISA, especially in the high-mass regime, where relative errors are below 
$1\%$ for every binary  configuration with $M\gtrsim 5\times 10^4M_\odot$. 

\subsection{Testing GR}

Our results suggest that, in generic theories of gravity, the GW signal from a BH coalescence contains a 5PN term which depends on the TLNs. Although the inclusion of tidal corrections in BH binaries is important for a correct modeling of the waveform, in most theories this term is subleading relative to other corrections coming both from dissipative effects (for example due to scalar- or vector-wave emission) and from corrections to the Hamiltonian of the binary~\cite{Yunes:2013dva,Berti:2015itd}.  
Nonetheless, there are cases in which the tidal deformability found in this work is the dominant correction relative to the GR waveform. This is the case of Chern-Simons gravity, where corrections to the GW phase enter at 2PN order if the components of the binary are spinning~\cite{Yagi:2012vf}, but at higher order in the absence of spin. Indeed, tidal heating of the horizon enters at 5PN order~\cite{Yagi:2011xp} whereas the leading point-particle Chern-Simons correction in the nonspinning case enters at 7PN~\cite{Yagi:2011xp,Pani:2011xj}. Since the quadrupolar axial TLN, $k_2^B$, enters at 6PN order in the waveform~\cite{Yagi:2013sva}, for nonspinning binaries the tidal correction will be subleading relative to tidal heating, but dominant relative to the point-particle phase. In this case, it would be interesting to estimate to which level the Chern-Simons parameter can be constrained by GW observations of BH binaries.

Unfortunately, the polar TLNs of Schwarzschild BHs in Chern-Simons gravity are zero as in GR and the axial TLNs enter the waveform only to subleading order~\cite{Yagi:2013sva}. Because $k_2^B\sim \zeta_{\rm CS}^2$ [cf.\ Eq.~\eqref{DCSl2}], in order to place an upper bound on the Chern-Simons coupling at the level of (say) $\zeta_{\rm CS}\lesssim 0.2$, one should be able to constrain the quadrupolar axial TLN at the level of $k_2^B\lesssim0.05$. 
The results shown in Fig.~\ref{fig:detectabilitygeneric} show that this level of precision cannot even be reached for the polar TLN $k_2^E$ with ET. Because the axial TLN affect the waveform at higher PN order~\cite{Yagi:2013sva}, we expect that it would be even more challenging to constrain them. 
On the other hand, the constraints on the dimensionless coupling $\zeta_{\rm CS}$ will significantly improve with LISA, but they will translate in poor bounds on the dimensionful coupling of the theory, $\alpha_{\rm CS}:=\zeta_{\rm CS} M^2$, due to the large mass of LISA sources.

Finally, for spinning BH binaries in Chern-Simons gravity, the leading correction enters at 2PN order and it is quadratic in the spin~\cite{Yagi:2011xp}. Because the first contribution to the TLN in Chern-Simons gravity comes from the quadrupolar, axial TLN, $k^B_2$, which enters at 6PN order~\cite{Yagi:2013sva}, our results should provide a reliable estimate for binaries with dimensionless spin $s\ll v_{\rm orb}^4$, where $v_{\rm orb}$ is the typical orbital velocity corresponding to the peak of sensitivity in the detector band.
Interestingly, for spinning BHs in Chern-Simons gravity, we expect that also the polar TLNs will be nonzero at linear order in the spin due to the spin-tidal effects discussed in Refs.~\cite{Poisson:2014gka,Pani:2015hfa,Landry:2015zfa,Pani:2015nua}. In this case, the tidal effects on the waveform might be stronger than the 2PN quadratic-in-spin corrections for moderately spinning BHs, albeit subleading in terms of PN order counting relative to dissipative effects.

\section{Discussion and extensions} \label{sec:conclusion}
The theory of the tidal deformability of compact objects has attracted considerable attention over the last few years. So far, applications of this theory have been mostly limited to astrophysics and to the possibility of constraining the equation of state of NSs with GW observations.
In this paper, we argued (implementing, through specific models, previous ideas and proposals~\cite{Porto:2016pyg,Porto:2016zng,Rothstein:simons}) that tidal effects can also be used to explore fundamental questions related to the nature of event horizons, the existence of ECOs, and the behavior of gravity in the strong-field regime.

Our main results can be summarized as follows:
\begin{itemize}
 \item In the framework of GR, the TLNs of ECOs are generically nonzero. In the limit that the ECO compactness $C\to 1/2$, all TLNs vanish but only logarithmically. This result holds for all models of ECO we have considered. It is therefore natural to conjecture that this logarithmic dependence is a generic feature of ultracompact exotic objects.\footnote{Of course, we are implicitly assuming that the radial coordinate $r$, in terms of which the compactness is defined, is special. Nevertheless, some models do suggest corrections at the Planck scale {\it in this same radial coordinate}, and it arises naturally also when expressing the results in terms of proper length~\cite{Holdom:2016nek}.} 
 \item The TLNs of a charged BH in Einstein-Maxwell theory and of an uncharged static BH in Brans-Dicke theory vanish, as in GR. 
These are both compelling extensions of vacuum GR, but our results indicate that the TLNs of BHs are nonzero in other interesting extensions. In particular, we have explicitly shown that the axial TLNs of a Schwarzschild BH in Chern-Simons gravity are nonzero. This result also extends to the TLNs of static BHs in Einstein-dilaton-Gauss-Bonnet gravity~\cite{Berti:2015itd}, but a full analysis is left for future work~\cite{inprep}. In this case, the presence of extra charges makes the multipolar expansion more involved. Preliminary investigation indicates that both the polar and the axial TLNs of static BHs in Einstein-dilaton-Gauss-Bonnet gravity are proportional to the coupling constant squared, similarly to the Chern-Simons case [cf.\ Eq.~\eqref{DCSl2}]. 
 \item We have explored the detectability of these tidal effects in some details, both for ground- and for space-based detectors. Ground-based detectors such as AdLIGO and ET can constrain ECO models with compactness\footnote{It is also interesting to note that the bounds on the compactness derived from the measurement of the TLNs with AdLIGO (respectively, ET) are slightly less (respectively, more) stringent than those that can be indirectly derived by the merger frequency and individual masses of GW150914~\cite{Abbott:2016blz}, which suggest a lower limit on the compactness of the two bodies, $C\gtrsim0.25$.} $C\lesssim0.2$ and $C\lesssim0.35$, respectively, whereas a LISA-like mission can constrain supermassive ECOs up to $C\lesssim0.49$. Interestingly, Advanced LIGO can set stringent constraints on various BS models, and both ET and LISA will be able to discriminate a BS binary from a BH binary just by measuring the TLNs of the binary components.

 \item The prospects for testing deviations from GR are less promising. While the TLNs of BHs beyond GR are different from zero, their effect in the GW signal is small and typically subleading relative to other, point-particle, beyond-GR effects such as dipolar emission. Nevertheless, the nonvanishing of the BH TLNs remains a smoking gun of deviations from GR and its phenomenological implications are under investigation~\cite{inprep}.
\end{itemize}

At the same time, our work is intended to be only a first step in understanding the tidal deformability of ECOs and of BHs beyond GR; as such, it can be extended in several interesting ways:
\begin{itemize}
 \item We neglected the presence of tidal fields of different nature, for example EM tidal fields in Einstein-Maxwell theory or scalar fields in Brans-Dicke theory or Chern-Simon gravity. We anticipate that the presence of extra tidal fields will give rise to new families of TLNs, which are related to the mass/current multipole moments induced by a (scalar or vector) extra tidal field~\cite{inprep}.
 \item We focused on nonrotating objects. The spin of the individual components of a neutron-star binary are typically small, but this might not be the case for ECOs and BHs. In general, subleading spin effects might be included by applying the formalism developed in Refs.~\cite{Poisson:2014gka,Pani:2015hfa,Landry:2015zfa,Pani:2015nua} to the systems studied in this work.
 \item We have estimated the detectability of tidal corrections in the GW signal from BH binaries beyond GR only in the simple case in which the leading-order correction to the GR waveform enters at 5PN order. A natural extension of our work is to include tidal corrections also when they are subleading relative to other, point-particle, beyond-GR terms. A related point is the fact that, when the TLNs are small, their effect might be smaller than other point-particle terms entering the waveform at 5PN order. In this case, the full 5PN waveform might be needed in order to extract the TLNs properly.\footnote{For NSs, the TLNs are enhanced by a factor $(R/M)^5\sim 10^5$, alleviating this issue; we are indebted to Leonardo Gualtieri for highlighting this point.}
\end{itemize}
%

\begin{acknowledgments}
We thank Leonardo Gualtieri for interesting discussions and Emanuele Berti and Gergios Pappas for helpful correspondence.
V. C. and P. P. gratefully acknowledge support from the Simons Center for Geometry and Physics, Stony Brook University at which some of the research for this paper was performed.
V.C. acknowledges financial support provided under the European Union's H2020 ERC Consolidator Grant ``Matter and strong-field gravity: New frontiers in Einstein's theory'' grant agreement no. MaGRaTh--646597. Research at Perimeter Institute is supported by the Government of Canada through Industry Canada and by the Province of Ontario through the Ministry of Economic Development $\&$
Innovation.
This project has received funding from the European Union's Horizon 2020 research and innovation programme under the Marie Sklodowska-Curie grant agreement No 690904, the ``NewCompstar'' COST action MP1304, and from FCT-Portugal through the project IF/00293/2013.
The authors thankfully acknowledge the computer resources, technical expertise and assistance provided by CENTRA/IST. Computations were performed at the clusters
``Baltasar-Sete-S\'ois'' and Marenostrum, and supported by the MaGRaTh--646597 ERC Consolidator Grant.
\end{acknowledgments}

\appendix
\section{TLNs of neutron stars\label{section:TLN_stars}}
It is interesting to compare the TLNs of an ECO with those for a NS, in order to investigate whether GW measurements of the tidal deformability can be used to distinguish an ECO from an ordinary compact star. For completeness, in Table~\ref{tab:EOS} we report some fitting formulas for the polar and axial TLNs of a NS with two different equations of state. The order of magnitude of the TLNs for compact NSs is given in Table~\ref{tab:summary}.

\begin{table*}[ht]
\centering
\caption{\footnotesize Coefficients of the fit $k_l^{E}\sim C^{-(2l+1)}\sum_{i=0}^4 a_i (1/2-C)^i$ and $k_l^{B}\sim C^{-2l}\sum_{i=0}^4 a_i (1/2-C)^i$ for the TLNs of a static neutron star with a stiff (\texttt{MS1}~\cite{Mueller:1996pm}) and relatively softer (\texttt{SLy4}~\cite{Douchin:2001sv}) equation of state. Data are taken from Ref.~\cite{Pani:2015hfa} and agree with the results in Ref.~\cite{Binnington:2009bb} after using the conversion in Eq.~\eqref{conversion}.}
\label{tab:EOS}
\begin{tabular}{|c|cccccc|} 
\hline
\hline
 EOS 	      & 		& $a_0$ & $a_1$ & $a_2$ & $a_3$ & $a_4$  \\
 \hline
 \multirow{4}{*}{\texttt{MS1}} & $k_2^E$  & -0.581& 8.721& -48.69& 123.4& -112.8  \\
			       & $k_3^E$  & -0.207& 3.230& -18.66& 47.82& -43.87 \\
			       & $k_2^B$  & -0.096& 1.369& -7.099& 17.53& -16.11 \\
			       & $k_3^B$  & -0.034& 0.514& -2.910& 7.511& -7.009 \\			     
  \hline
 \multirow{4}{*}{\texttt{SLy4}} & $k_2^E$  & -0.414& 6.227& -35.35& 92.70& -87.70  \\
				& $k_3^E$  & -0.150& 2.326& -13.57& 35.56& -33.38 \\
				& $k_2^B$  & -0.063& 0.876& -4.483& 11.57& -11.24 \\
				& $k_3^B$  & -0.023& 0.346& -1.978& 5.288& -5.115 \\						
\hline
\hline
\end{tabular}
\end{table*}
%

\section{Determination of TLNs}\label{app:comp}
In order to compute the TLNs we need to calculate the expressions for the induced mass and current multipole moments as a function of the external tidal field. We use a linear perturbation theory approach to disturb slightly the spacetime metric,
\begin{equation}
g_{\mu\nu}=g_{\mu\nu}^{^{(0)}}+h_{\mu\nu}\,,\label{metricpertdef}
\end{equation}
where $g_{\mu\nu}^{^{(0)}}$ is the background spacetime metric and $h_{\mu\nu}$ is a small perturbation. We focus on static, spherically symmetric background metrics which are described by
%
\begin{equation}
\label{staticmetric}
g^{^{(0)}}_{\mu\nu}={\rm diag}\left(-e^{\Gamma},e^{\Lambda_g},r^2,r^2\sin^2{\!\theta}\,\right).
\end{equation}
We decompose $h_{\mu\nu}$ in spherical harmonics and separate the perturbation in even and odd parts, $h_{\mu\nu}=h_{\mu\nu}^{\text{even}}+h_{\mu\nu}^{\text{odd}}$, according to parity. In the Regge-Wheeler gauge~\cite{Regge:1957td}, $h_{\mu\nu}$ can be decomposed as
\begin{widetext}
\begin{align}
\label{heven}
&h_{\mu\nu}^{\text{even}}=
\left(\begin{array}{cccc}
e^{\Gamma}H_0^{lm}(t,r)Y^{lm} & H_1^{lm}(t,r)Y^{lm} & 0 & 0 \\
 H_1^{lm}(t,r)Y^{lm} & e^{\Lambda_g}H_2^{lm}(t,r)Y^{lm} & 0 &0\\
 0 &0 & r^2 K^{lm}(t,r)Y^{lm}& 0\\ 
0&0 &0 & r^2\sin^2{\theta} K^{lm}(t,r)Y^{lm}
\end{array}\right)
,\\
\nonumber\\
\label{hodd}
&h_{\mu\nu}^{\text{odd}}=\left(\begin{array}{cccc}
0 & 0 & h_0^{lm}(t,r)S_\theta^{lm} & h_0^{lm}(t,r)S_\varphi^{lm} \\
0 &0 & h_1^{lm}(t,r)S_\theta^{lm} & h_1^{lm}(t,r)S_\varphi^{lm} \\
h_0^{lm}(t,r)S_\theta^{lm} &h_1^{lm}(t,r)S_\theta^{lm}  &0 &0\\
h_0^{lm}(t,r)S_\varphi^{lm}& h_1^{lm}(t,r)S_\varphi^{lm} & 0 &0 \end{array}
\right),
\end{align}
\end{widetext}
with $\left(S_\theta^{lm},S_\varphi^{lm}\right)\equiv \left(-Y_{,\varphi}^{lm}/\sin{\theta},\,\sin{\theta} \,Y_{,\theta}^{lm}\right).$	

In the presence of scalars or vectors, spacetime fluctuations are accompanied by the corresponding fluctuations in these fields,
\begin{align}
\label{4pot}
&A_\mu=A^{^{(0)}}_\mu+\delta A_\mu,\\
\label{scalarfield}
&\Phi=\Phi^{^{(0)}}+\delta\Phi,
\end{align}
where $A^{^{(0)}}$ and $\Phi^{^{(0)}}$ are background quantities while $\delta A_\mu$ and $\delta\Phi$ are small perturbations. We expand the scalar perturbation $\delta\Phi$ and $4$-potential $\delta A_{\mu}$ as~\cite{Rosa:2011my,Pani:2013wsa},
\begin{eqnarray}
\delta\Phi &=&\delta\phi_{lm}Y^{lm}\,,\label{eq:scalardecomp}\\
\delta A_{\mu}&=&\left(\frac{u_1^{lm}}{r} Y^{lm},\frac{u_2^{lm}\,e^{-\Gamma}}{r} Y^{lm} ,\frac{u_3^{lm} Y_{b}^{lm}+u_4^{lm} S_{b}^{lm}}{{\eta}}\right)\,,\nn\\\label{eq:EMpot}
\end{eqnarray}
with \mbox{$Y_{b}^{lm}\equiv\left(Y_{,\theta}^{lm},Y_{,\varphi}^{lm}\right)$} and ${\eta}=l(l+1)$.
Hereafter, we shall drop the $(lm)$ superscripts on all quantities with the exception of multipole moments.

Expressions for the metric functions in~\eqref{heven} and~\eqref{hodd}, the electromagnetic functions in~\eqref{eq:EMpot}, and the scalar fields in~\eqref{eq:scalardecomp} can be obtained by solving the linearized field equations for a given model. The remaining task consists in extracting the multipole moments and tidal fields from the spacetime metric. Thorne developed a method to define the multipole coefficients of any spacetime metric given in coordinates which are asymptotically Cartesian and mass centered (ACMC)~\cite{Thorne:1980ru}. Another definition of the multipole moments of an axisymmetric and asymptotically flat spacetime was given by Geroch and Hansen~\cite{Geroch:1970cd,Hansen:1974zz}. These two distinct definitions of moments were shown to be equivalent~\cite{Gursel:1983}.

The multipole moments can be extracted from the asymptotic behavior of the spacetime metric and fields,
\begin{widetext}
\begin{align}
\label{eq:gttexpansion}
g_{tt}&=-1+\frac{2M}{r}+ \sum_{l\geq 2}\left(\frac{2}{r^{l+1}}\left[\sqrt{\frac{4\pi}{2l+1}}M_l Y^{l0}  + (l'<l\,\text{pole})\right]-\frac{2}{l(l-1)}r^l\left[\mathcal{E}_l Y^{l0}+(l'<l\,\text{pole})\right]\right),\\
\label{eq:gtphiexpansion}
g_{t\varphi}&=\frac{2J}{r}\sin^2{\theta}+\sum_{l\geq 2}\left( \frac{2}{r^l}\left[\sqrt{\frac{4\pi}{2l+1}}\frac{S_l}{l}S_\varphi^{l0}+\left(l'< l\,\text{pole}\right)\right]+\frac{2r^{l+1}}{3l\left(l-1\right)}\left[\mathcal{B}_l S^{l0}_\varphi+\left(l'< l\,\text{pole}\right)\right] \right),\\
\label{eq:Atexpansion}
A_t&=-\frac{Q}{r}+\sum_{l\geq 1}\left(\frac{2}{r^{l+1}}\left[\sqrt{\frac{4\pi}{2l+1}}Q_l Y^{l0} +\left(l'<l\,\text{pole}\right)\right]-\frac{2}{l(l-1)}r^l\left[\mathfrak{E}_lY^{l0}+\left(l'< l\,\text{pole}\right)\right]\right),\\
\label{eq:Aphiexpansion}
A_\varphi &=\sum_ {l\geq 1} \left(\frac{2}{r^l}\left[\sqrt{\frac{4\pi}{2l+1}}\frac{J_l}{l} S_\varphi^{l0} + \left(l'<l\,\text{pole}\right)\right]+\frac{2r^{l+1}}{3l\left(l-1\right)}\left[\mathfrak{B}_l  S_\varphi^{l0}+\left(l'< l\,\text{pole}\right)\right]\right),\\
\label{eq:scalarexpansion}
\Phi&=\Phi_0 +\sum_{l\geq 1} \left(\frac{1}{r^{l+1}}\left[\Phi_l Y^{l0} + (l'<l \,\text{pole})\right]- r^l \left[\mathcal{E}^\text{S}_l+\left(l'< l\,\text{pole}\right)\right]\right),
\end{align}
\end{widetext}
where we defined $Q_l$, $J_l$, $\Phi_l$ as the electric, magnetic and scalar multipole moments, respectively, and $\mathfrak{E}_l$, $\mathfrak{B}_l$ and $\mathcal{E}^\text{S}_l$ as the electric, magnetic and scalar tidal field moments, respectively.

The decomposition of the scalar field and the vector potentials were chosen so they could be easily compared with the metric decomposition.
An appropriate comparison between the solution of the field equations and the expansions~\eqref{eq:gttexpansion} to \eqref{eq:scalarexpansion} allows us to extract the multipole moments and, in turn, the TLNs as defined in Eq.~\eqref{Lovenumbersdef1}.

\section{Tidal perturbations of boson stars}\label{app:BS}
%
\begin{figure*}[th]
\includegraphics[width=0.45\textwidth]{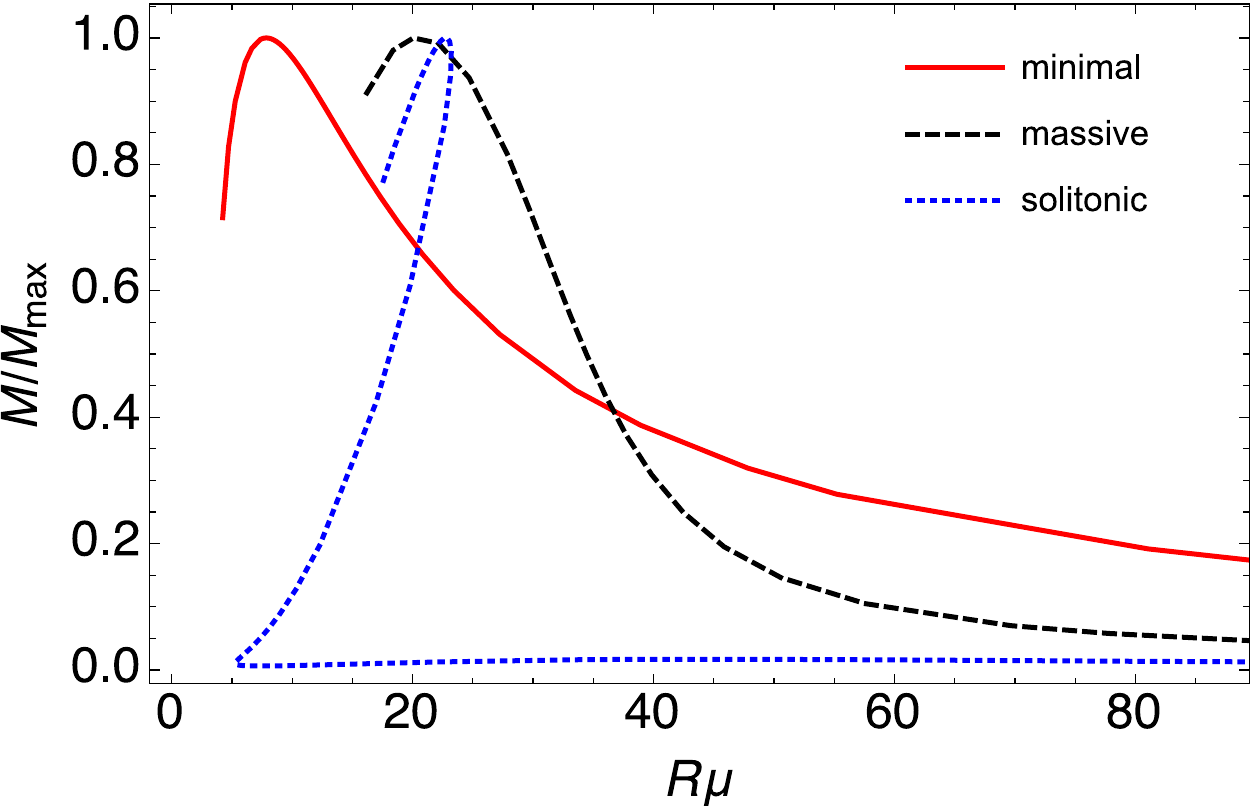}\quad
\includegraphics[width=0.45\textwidth]{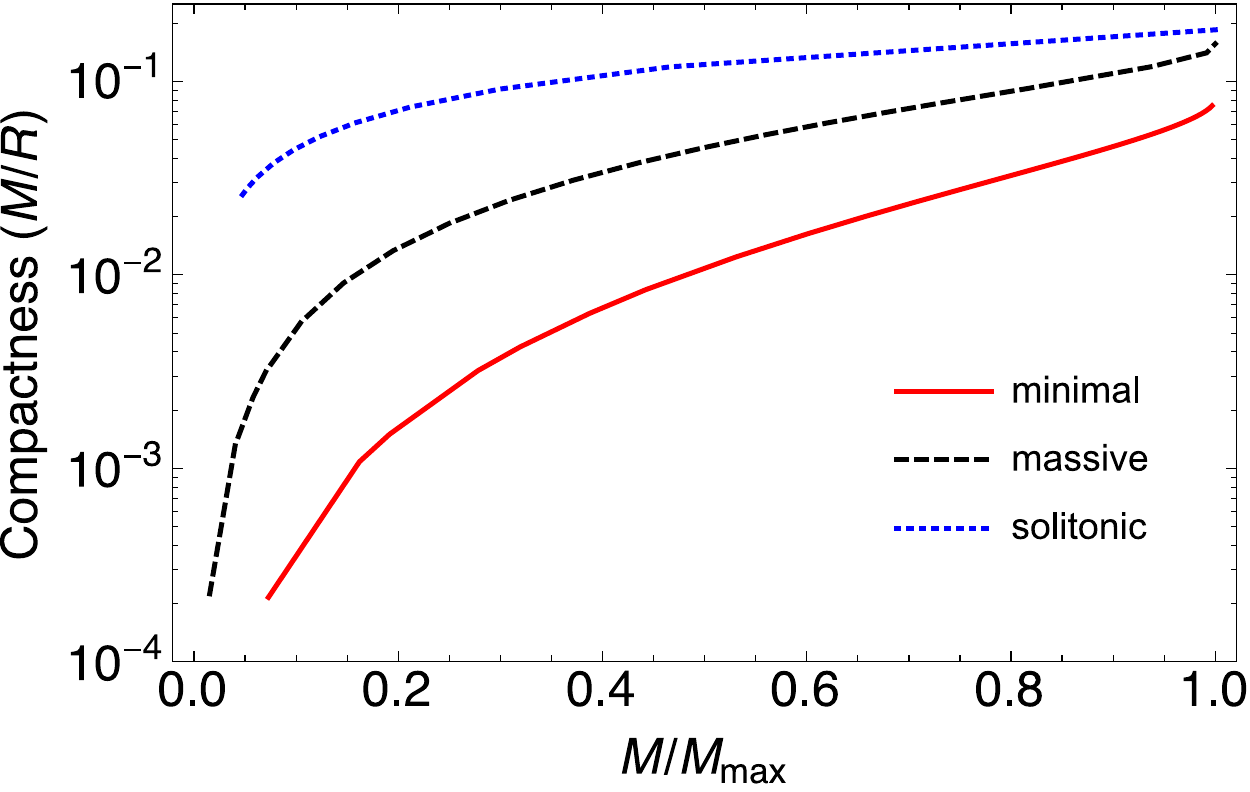}
\caption{Left panel: ADM mass as a function of the effective radius $R$ for different models of BSs,
including some unstable configurations (to the left of maximum mass).
Right panel: Compactness of the models in the stable branch as a function of the mass.
For massive and solitonic BSs we have considered $\alpha=10^4\mu^2$ and $\sigma_0=0.05$, respectively.}\label{fig:massradius}
\end{figure*}
%

\subsection{Background solutions}

We consider spherically symmetric BSs, with background metric given by Eq.~\eqref{staticmetric},
and the following ansatz for the background scalar field,
\be
\Phi^{^{(0)}}(t,r) = \phi_0(r)\,e^{-i\omega t}.
\ee
Despite the time dependence of the scalar field, the Einstein-Klein-Gordon background equations are static,
\beq
\Lambda_g' &=& \frac{1-e^{\Lambda_g}}{r} + 8\pi r \left(\omega^2 e^{\Lambda_g-\Gamma}\phi_0^2+\phi_0'^2 + e^{\Lambda_g} V\right),\label{ekg1}\\
\Gamma' &=& \frac{e^{\Lambda_g}-1}{r} + 8\pi r \left(\omega^2 e^{\Lambda_g-\Gamma}\phi_0^2+\phi_0'^2 - e^{\Lambda_g} V\right),\label{ekg2}\\
\phi_0'' &=& e^{\Lambda_g}\left(\frac{d V}{d|\phi^2|}-\omega^2 e^{-\Gamma}\right)\phi_0 + \left(\frac{\Lambda_g'-\Gamma'}{2}-\frac{2}{r}\right)\phi_0'\,, \nn\\\label{ekg3}
\eeq

The equilibrium spherically symmetric BS solutions are found by integrating numerically
Eqs.~\eqref{ekg1} to \eqref{ekg3} along with suitable boundary conditions. Namely, we impose regularity at the
origin, $\Gamma=\Gamma_c$, $\Lambda_g=0$, $\phi=\phi_c$, $\phi'=0$,
whereas at infinity we impose the metric to be Minkowski and the scalar field to vanish.
For a given value $\phi_c$ of the scalar at the center of the star,
the problem is then reduced to an eigenvalue problem for the frequency $\omega$,
which we solve through a standard shooting method. The value of $\Gamma_c$ is arbitrary
and can be tuned in order to have $\Gamma(r\to\infty)=0$.

The total mass of the solution is $M=m(r\to\infty)$, where $m(r)$ is defined by
\be
e^{-\Lambda_g(r)}\equiv1-\frac{2m(r)}{r}\,.
\ee

Contrary to the minimal and massive case --~ in which the scalar profile decays exponentially~-- in the solitonic model the scalar profile has a very steep profile which makes the numerical integration of the background equations very challenging, requiring very fine-tuned shooting parameters~\cite{Macedo:2013jja}.

BSs do not have a hard surface,
as the scalar is spread out all over the radial direction.
However, the configuration is highly localized in a radius $\sim1/\mu$ and it
is customary to define the effective radius $R$ as the radius
within which the 99\% of the total mass is contained, i.e.\ $m(R)=0.99M$.
This is slightly different for the solitonic model, in this case the steepness of
the scalar profile makes the definition of the radius more natural.

By using the above procedure, we can compute a sequence of background solutions
characterized by the value of the scalar at the center
of the star, $\phi_c$. In Fig.~\ref{fig:massradius}, we show the gravitational mass $M$ against $R$ (left panel) and the compactness $C=M/R$ as a function of the mass (right panel).
For minimal and interacting BSs, we observe a maximum which separates stable (on the right)
from unstable (on the left) configurations~\cite{Gleiser:1988rq,Gleiser:1988ih,Hawley:2000dt}.
On the other hand, solitonic BSs display a completely different behavior: there exists a stable branch
for high values of $R$ and small masses (bottom-right part of the left panel in Fig.~\ref{fig:massradius}),
an unstable branch that starts after the first maximum roughly at $R\mu\sim50$, and
then a second stable branch which starts at $R\mu\sim10$ up to the maximum
on the top-right part of the plot.

As shown in the right panel of Fig.~\ref{fig:massradius}, for minimal BSs the compactness is typically of ${\cal O}(0.01)$, and slightly larger for massive BSs.
On the other hand, solitonic BSs can be almost as compact
as BH (i.e., $C\approx1/2$), meaning that their radius can be of order of
the Schwarzschild light ring~\cite{Cardoso:2014sna,Cardoso:2016oxy}.

\subsection{Perturbations and TLNs}

\subsubsection{Polar perturbations}
We consider perturbations of the equilibrium configuration, sourced by an external static tidal field.
The metric perturbation is given by~\eqref{heven} with time independent functions.
We write the scalar perturbation as 
\be\label{phiperturb}
\delta\Phi = \sum_{l,m} e^{-i\omega t}\phi_1(r)Y^{lm}(\theta,\varphi).
\ee

Plugging~\eqref{heven}
and~\eqref{phiperturb} into the linearized Einstein equations, we find $H_0=H_2\equiv H$ and $H_1=0$,
whereas $K$ can be written as a function of $H$, $\phi_1$ and of the background functions.
We are then left with a radial equation for the perturbation function $H$, which is coupled to the perturbed Klein-Gordon equation for the scalar perturbation $\phi_1$,
\begin{widetext}
\beq
H'' &&+ \left(\frac{2}{r}-8\pi r \omega^2 \phi_0^2 e^{\Lambda_g-\Gamma} + \Gamma' - 8\pi r \phi_0'^2\right) H'
+32\pi \left[\phi_0'' - \left(\frac{\Lambda_g'+\Gamma'}{2} - \frac{2}{r}\right)\phi_0' - \omega^2 \phi_0 e^{\Lambda_g-\Gamma}\right] \phi_1 \nn\\
&&+\left[\Gamma'^2-\frac{2\Gamma'}{r}-48\pi e^{\Lambda_g-\Gamma}\omega^2\phi_0^2 + 16\pi\phi_0'^2 + \frac{e^{\Lambda_g}(l^2+l+2)-2}{r^2}\right] H = 0\,,\label{eqH}\\
\phi_1'' &&+ \left(\frac{2}{r} - 8\pi  r \omega^2\phi_0^2 e^{\Lambda_g-\Gamma}+\Gamma'-8\pi  r\phi_0'^2\right)\phi_1'
+ \left[\phi_0'' - \left(\frac{\Lambda_g'+\Gamma'}{2} - \frac{2}{r}\right)\phi_0' - \omega^2 \phi_0 e^{\Lambda_g-\Gamma}\right] H\nn\\
&&-\left[\frac{\phi_0''}{\phi_0} + \left(\frac{\Gamma'-\Lambda_g'}{2} + \frac{2}{r}\right)\frac{\phi_0'}{\phi_0} + 32\pi\phi_0'^2 + \frac{l(l+1) e^{\Lambda_g}}{r^2}\right]\phi_1  - \frac{d \delta V}{d|\phi^2|} e^{\Lambda_g}\phi_0=0\,,\label{eqphi1}
\eeq
\end{widetext}
where $\delta V$ is the linear correction to the scalar potential, i.e. $\delta V\approx V(|\phi|^2)-V(|\phi_0|^2)$.
We now solve the perturbation system 
supplied by regular boundary conditions at the origin,
\be
H_0 \approx H_0^{(l)} r^l + \mathcal{O}\left(r^{l+2}\right),\quad
\phi_1 \approx \phi_1^{(l)} r^l + \mathcal{O}\left(r^{l+2}\right).
\ee

Since the system is linear, the value of $H_0^{(l)}$ can be set to 1,
and the correct value can be recovered a posteriori. The value of
$\phi_1^{(l)}$ is determined by requiring that $\phi_1\to0$ as $r\to\infty$
using a shooting method.

At distances $R_\text{ext}$ much larger than the BS effective radius, Eq.~\eqref{eqH} reduces to
\be\label{HBS}
H'' + \frac{2(r-M)}{r(r-2M)} H' -\frac{4 M^2 -2\eta M r + \eta r^2}{r^2 (r-2 M)^2} H = 0\,,
\ee
with ${\eta}:=l(l+1)$. Equation~\eqref{HBS} has a general solution in terms of the associate Legendre functions $P^2_l$ and $Q^2_l$.
Using their asymptotic behavior and comparing with~\eqref{eq:gttexpansion} we find~\cite{Hinderer:2007mb}
\begin{widetext}
\beq
k_2^E &=& \frac{8}{5}\left({(1-2\C)}^2 [2\C (y-1)-y+2]\right)
\left( 3{(1-2\C)}^2 [2\C (y-1)-y+2] \log\left({1-2\C}\right)\right.\nonumber\\
&&\left.\vphantom{{(1-2\C)}^2} + 2\C\left[4\C^4 (y+1) + 2\C^3 (3y-3) + 2\C^2 (13-11y) + 3\C (5y-8) - 3y+6\right]\right)^{-1}\,,\\
k_3^E &=& \frac{6}{7} \left({(1-2\C)}^2 \left[2\C^2 (y-1)-3\C (y-2)+y-3\right]\right)
\left(15 {(1-2\C)}^2 \left(2\C^2 (y-1)-3 \C (y-2)+y-3\right) \log (1-2\C)\right.\nonumber\\
&&\left.\vphantom{{(1-2\C)}^2} + 2\C \left[4\C^5 (y+1)+2\C^4 (9 y-2)-20\C^3 (7 y-9)+5\C^2 (37 y-72)-45\C (2 y-5)+15 (y-3)\right]\right)^{-1}.\qquad\qquad
\eeq
\end{widetext}
for $l=2$ and $l=3$, respectively, and where $\C=M/R_\text{ext}$ and $y=rH'/H$ evaluated at $R_\text{ext}$.
The values of $k_l^E$ are independent of the extraction radius $R_\text{ext}$ if the latter
is sufficiently large.

\subsubsection{Axial perturbations}
In this case, the metric perturbation is given by~\eqref{hodd}, while the scalar perturbation
is given again by~\eqref{phiperturb}.
The perturbed Einstein equations require $h_1=\phi_1=0$, and we are left with a single
radial equation for the perturbed function $h_0$,
\be
h_0'' - \frac{\Lambda_g'+\Gamma'}{2}\,h_0' 
+\frac{r (\Lambda_g'+\Gamma') - (l^2+l-2) e^{\Lambda_g}-2}{r^2}\,h_0 = 0\,,\label{eqh0}
\ee
that we solve along with regular boundary conditions at the origin,
\be
h_0\approx h_0^{(l+1)} r^{l+1} + \mathcal{O}\left(r^{l+3}\right).
\ee
Notice that the value of $h_0^{(l+1)}$ is not given but it can be fixed arbitrarily to 1
and corrected a posteriori once the intensity of the tidal field is known.

Outside the star, the equation for the odd perturbation~\eqref{eqh0} reduces to the simple differential equation
\be%
h_0'' + \frac{4M - l(l+1)r}{r^2 (r-2M)}\,h_0 = 0,
\ee%
whose solution can be written in terms of elementary functions once $l$ is fixed.
By matching its asymptotic behavior to~\eqref{eq:gtphiexpansion}, we find  (for $l=2$ and $l=3$)
\begin{widetext}
\be%
k_2^B = \frac{8}{5}\frac{2\C (y-2)-y+3}{2\C \left[2\C^3 (y+1) + 2\C^2 y + 3\C (y-1)-3 y+9\right] + 3[2\C (y-2)-y+3] \log (1-2\C)},
\ee%
\beq
k_3^B = \frac{8}{7} \left(\vphantom{{(1)}^2} 8\C^2 (y-2)-10\C (y-3)+3 (y-4)\right)
\left(\vphantom{{(1)}^2} 15\left[8 \C^2 (y-2)-10 \C (y-3)+3 (y-4)\right] \log (1-2 \C)\right.\nonumber\\
\left.\vphantom{{(1)}^2} + 2\C \left[4 \C^4 (y+1)+10 \C^3 y+30 \C^2 (y-1)-15 \C (7 y-18)+45 (y-4)\right]\right)^{-1},\qquad\qquad\qquad\quad
\eeq
\end{widetext}
where again $\C=M/R_\text{ext}$ but now $y=rh_0'/h_0$ evaluated at $R_\text{ext}$.
Even in this case, the values of $k_l^B$ are independent of the extraction radius $R_\text{ext}$ if the latter
is sufficiently large.

\section{Tidal perturbations of BH-like ECOs}\label{app:ECOs}

The exterior spacetime of all models considered in Sec.~\ref{sec:QC} is commonly described by the Schwarzschild metric. Therefore, the perturbation formalism is identical to the one developed for a nonrotating uncharged BH, where the metric is perturbed according to Eq.~\eqref{metricpertdef} and the even and odd sector perturbations can be described as Eqs.~\eqref{heven} and \eqref{hodd}. On the other hand, the interior and the junction/boundary conditions at the radius $r_0$ are model dependent. In this Appendix we discuss the procedure to compute the TLNs for these objects.

\subsection{Polar-type TLNs of BH-like ECOs}

\subsubsection{Exterior spacetime}
Let us first consider the exterior spacetime. Einstein's equations for static polar-type perturbations of the Schwarzschild metric lead to (the notation follows Appendix~\ref{app:comp})
\beq
&&\frac{d^2H_0}{dr_*^2}+\frac{2f}{r}\frac{dH_0}{dr_*}-\left(f\frac{l(l+1)}{r^2}+\frac{4M^2}{r^4}\right)H_0=0\,,\\
&&\frac{dK}{dr}=-\frac{f-1}{r f}H_0+H_0'\,,\\
&&K=\frac{[1+f(l^2+l-2-f)]H_0-rf(f-1)H_0'}{f(l^2+l-2)}\,. \label{Kext}
\eeq
Here $f=1-2M/r$ and primes stand for derivatives with respect to $r$. The above equations can be solved for~\cite{Hinderer:2007mb,Binnington:2009bb}
\begin{equation}
 H_0^{\rm ext}=C_1P_l^2(r/M-1)+C_2Q_l^2(r/M-1)\,, \label{H0ext}\\
\end{equation}
for any value of $l$, and where $C_1$ and $C_2$ are two integration constants. The term proportional to $C_1$ diverges at large distances and is identified with the external tidal field, whereas the term proportional to $C_2$ is the body's response. The metric function $K$ follows straightforwardly from Eq.~\eqref{Kext}.

\subsubsection{Interior spacetime}
The interior spacetime depends on the model under consideration. In the wormhole model, we consider that the other universe is an exact copy of exterior metric, so that polar perturbations are described by Eq.~\eqref{H0ext} with two independent constants, 
\begin{equation}
 H_0^{\rm int}=C_3P_l^2(r/M-1)+C_4Q_l^2(r/M-1)\,, \label{H0intWH}\\
\end{equation}
On the ``other side'' of the wormhole, we require that there are no tidal fields, i.e. $C_3=0$.

In the perfect-mirror model, perturbations do not penetrate the surface and the interior solution is irrelevant. On the other hand, in the gravastar model the interior solution which is regular at the origin reads 
\be\label{H0intGS}
H_0^{\rm int} \propto \frac{r_0^2 r^l}{r_0^2 - 2C r^2}\,
{}_2F_1\left(\frac{l-1}{2},\frac{l}{2};l+\frac{3}{2};\frac{2 C r^2}{r_0^2}\right)\,,
\ee%
for any value of $l$. 


\subsubsection{Matching conditions and TLNs}\label{matching}
In the wormhole and in the gravastar case the interior and the exterior solutions are overall described by three independent constants, whereas the perfect-mirror model is described by two constants. Since the problem is linear, an overall amplitude is irrelevant so we need to impose two junction conditions at $r=r_0$ in the wormhole and gravastar cases, and one boundary condition at $r=r_0$ in the perfect-mirror case.

In the former cases, we can impose the Darmois-Israel junction conditions~\cite{Israel:1966rt}, which relate the discontinuity of the extrinsic curvature across the radius with the properties of a thin shell of matter located at $r=r_0$. By adapting the formalism developed in Ref.~\cite{Pani:2009ss}, for polar perturbations, we find that $[[K]]=0$ and $[[dK/dr_*]]=-8\pi \sqrt{f(r_0)}\delta\Sigma$, where $\delta\Sigma$ is the perturbation of the surface energy density of the thin shell and the symbol ``$[[\ldots]]$'' denotes the ``jump'' of a given quantity across
the spherical shell, i.e. $[[A]] \equiv \lim_{\epsilon\to0}{A(r\to r_0 +\epsilon)-A(r\to r_0-\epsilon)}$. For simplicity, we assume that the thin-shell material is stiff, so that $\delta\Sigma\sim 0$. Therefore, in the polar sector we impose the following conditions
\begin{equation}
 [[K]]=0 \,, \qquad [[dK/dr_*]]=0\,. \label{junctionsPolar}
\end{equation}
These two conditions completely specify the matching between the interior and the exterior solution in the wormhole and gravastar cases. In the latter case, the junction conditions~\eqref{junctionsPolar} agree with those derived in Ref.~\cite{Uchikata:2016qku}.

In the perfect-mirror case, we shall impose a $Z_2$ symmetry on the surface and, therefore, the wave function vanishes at $r=r_0$. In this case, one can solve for the Zerilli function $\Psi_{\rm Z}$ in the static limit and then reconstruct the metric function $H_0$ through~\cite{Zerilli:1970se}
\begin{eqnarray}
 H_0 &=& \frac{1}{r (6 M+{\gamma_l}  r)} \left[ \left({\gamma_l}  r^2-6 M^2-3 {\gamma_l}  M r\right) \Psi_Z'(r)\right.\nn\\
 &&\left.+\frac{\Psi_Z(r)}{6 M+{\gamma_l}  r} \left(36 M^3/r+18 {\gamma_l}  M^2+3 {\gamma_l} ^2 M r\right.\right.\nn\\
 &&\left.\left.+{\gamma_l} ^2 ({\gamma_l}/2 +1) r^2\right)\right]\,, 
\end{eqnarray}
where ${\gamma_l}=(l-1)(l+2)$. In the static limit, $H_1=0$ and $K$ can be also written in terms of $\Psi_\text{Z}$ and of its derivative. The Zerilli function satisfies a second-order differential equation~\cite{Zerilli:1970se}, which can be solved for analytically in the static limit.
By imposing the Dirichlet boundary condition $\Psi_{\rm Z}=0$, the ratio $C_1/C_2$ in Eq.~\eqref{H0ext} is completely specified.

Finally, once the perturbation equations are completely specified (modulo an overall amplitude) through the junction/boundary conditions, it is straightforward to compare the analytical expression for $H_0$ at large distance with Eq.~\eqref{eq:gttexpansion}, extract the multipole moments and the tidal-field amplitudes, and finally compute the polar TLNs by using Eq.~\eqref{Lovenumbersdef1}.
Interestingly, this procedure yields exact results in closed form, although the general expression for the polar TLNs is cumbersome. The full result is provided online~\cite{webpage}, whereas the high-compactness regime is discussed in the main text.

\subsection{Axial-type TLNs of BH-like ECOs}

\subsubsection{Exterior spacetime}
Einstein's equations for static axial-type perturbations of the Schwarzschild metric imply $h_1=0$ and 
\begin{equation}
{\cal D}_A^{(l)}h_0 \equiv
h_0''+\frac{4 M-l (l+1) r}{r^2 (r-2 M)}\,h_0=0\,,
\end{equation}
which can be solved for $h_0$ in terms of hypergeometric functions. The general solution reads~\cite{Binnington:2009bb}
%
\begin{eqnarray}
h_0^{\rm ext} &=& c_1\,r^2 \, {}_2F_1\left(1-l,l+2;4;\frac{r}{2 M}\right)\nn\\
&+&c_2\,G_{2,2}^{2,0}\left(\frac{r}{2 M}\left|
\begin{array}{c}
 1-l,l+2 \\
 -1,2 \\
\end{array}\right.\right),\label{h0ext}
\end{eqnarray}
%
where $G_{2,2}^{2,0}$ is the Meijer function and ${}_2F_1$ is one of the hypergeometric functions. The terms proportional to $c_1$ and $c_2$ are identified with the external tidal field and with the body response, respectively. The above solution reduces to simple expressions for integer values of $l$, which can be written in terms of polynomial and logarithmic functions.

\subsubsection{Interior spacetime}

Also in the axial case, the interior spacetime is model dependent. In the wormhole case we consider the same solution as in Eq.~\eqref{h0ext} but with zero tidal field, namely
\begin{eqnarray}
h_0^{\rm int} &=& c_3\,G_{2,2}^{2,0}\left(\frac{r}{2 M}\left|
\begin{array}{c}
 1-l,l+2 \\
 -1,2 \\
\end{array}\right.\right). \label{h0intWH}
\end{eqnarray}

In the gravastar case, the interior solution which is regular near the origin reads
\be\label{h0intGS}
h_0(r) \propto r^{l+1} \, _2F_1\left(\frac{l-1}{2},\frac{l+2}{2};l+\frac{3}{2};\frac{2 C r^2}{r_0^2}\right)\,,
\ee%
for any value of $l$. Finally, as in the polar case, the interior solution of the perfect-mirror model is irrelevant for the purposes of computing the TLNs.

\subsubsection{Matching conditions and TLNs}
The junction conditions for axial perturbations are easier because they do not couple to the matter of a putative thin shell. In the dynamical case, they simply read $[[h_0]]=0=[[h_1]]$~\cite{Pani:2009ss}. In the static case, $h_1$ vanishes identically and one is left with a single second-order differential equation for $h_0$. Therefore, regularity of the axial perturbations across the shell imposes that $h_0$ and its derivative with respect to $r_*$ be smooth. Thus, for the wormhole and gravastar cases in the axial sector we impose
\begin{equation}
 [[h_0]]=0\,, \qquad [[d h_0/dr_*]]=0\,. \label{junctionAxial}
\end{equation}

For the perfect-mirror model, we follow the same procedure outlined in Appendix~\ref{matching}, namely we impose a Dirichlet condition on the Regge-Wheeler function $\Psi_{\rm RW}$ evaluated at $r=r_0$. This function is defined as\footnote{Note that our definition differs from the standard one by a factor $\omega$, which has been included so that $h_1\to0$ in the static limit, whereas $h_0$ remains finite, as expected.}
\begin{eqnarray}
h_0=\frac{d(r\Psi_{\rm RW})}{dr_*}\,,\qquad h_1=\frac{-i\omega r }{1-2M/r}\Psi_{\rm RW}\,, \label{defh0}
\end{eqnarray}
and satisfies the Regge-Wheeler equation~\cite{Regge:1957td}. The latter can be solved analytically in the static limit, $\omega=0$. Again, the ratio of the two integration constants in Eq.~\eqref{h0ext} is fixed by imposing $\Psi_{\rm RW}=0$. 

After the perturbations are fully specified through the junction/boundary conditions, the axial TLNs can be computed by comparing the large-distance behavior of $h_0$ with Eq.~\eqref{eq:gtphiexpansion}, extracting the multipole moments, and finally using the definition~\eqref{Lovenumbersdef1}.
As in the polar case, we find a closed-form, cumbersome expression for the axial TLN~\cite{webpage}, whose high-compactness limit is provided in the main text for the various models.

\bibliographystyle{apsrev4}
\bibliography{refs}
	
\end{document}